\documentclass[aps,a4paper]{elsarticle}
\usepackage{amsmath}
\usepackage{amsfonts}
\usepackage{amssymb}
\usepackage{graphics}
\usepackage{epsfig}
\usepackage{subfigure}
\usepackage{natbib}
\begin{document}
\title{Negative partial density of states in mesoscopic systems}
\author{Urbashi Satpathi and P. Singha Deo$ ^{\dagger} $}
\address{$ ^{\dagger} $deo@bose.res.in\\
Unit for Nano Science and Technology,
S. N. Bose National Centre for Basic Sciences, 
JD Block, Sector III, Salt Lake City, Kolkata 98, India.}
\date{\today}
\begin{abstract}
Since the experimental observation of quantum mechanical scattering phase shift in mesoscopic systems, several aspects of it has not yet been understood. The experimental observations has also accentuated many theoretical problems related to Friedel sum rule and negativity of partial density of states. We address these problems using the concepts of Argand diagram and Burgers circuit. We can prove the possibility of negative partial density of states in mesoscopic systems. Such a conclusive and general evidence cannot be given in one, two or three dimensions. We can show a general connection between phase drops and exactness of semi classical Friedel sum rule. We also show Argand diagram for a scattering matrix element can be of few classes based on their topology and all observations can be classified accordingly. 
\end{abstract}
\maketitle
\section{Introduction}
A number of experimental \cite{schu,yang,yang1,kob1,kob2,hei,zaf} and theoretical works \cite{deolegett,deo1,tan,eng,levy,hack,singlechannel,3prong,solis,vass} have studied the scattering phase shift in low dimensional mesoscopic systems where the electron dynamics is determined by quantum mechanics. Essentially, the asymptotic states are that of free particles in a quantum wire and the scatterer is embedded in the path of this quantum wire \cite{hack,ulrich}. Model specific approach so far has led to contradictory and confusing results \cite{zaf,tan,lee,levy,vass,buttiker2}. Heiblum et al \cite{zaf} quote that `There is by now vast theoretical evidence that the transmission phase depends on the specific properties of the QD's levels that participate in the transport'. However, in all these studies there is no reference to well established theorems and results on discontinuous phase changes and lapses known for a long time in the community that studies classical wave trains \cite{mic2, mic1}. Such phase changes are still very important in mesoscopic physics as they are related to breakdown of parity effect \cite{deolegett}, interpretation of Friedel sum rule (FSR) \cite{tan,levy,singlechannel,deo2}, relation to partial density of states (PDOS) \cite{buttiker2,gsp,buttiker3,buttiker4,buttiker1,n1}, etc. that determine the thermodynamic properties of a mesoscopic system which is absent in case of classical waves. So in this work we first recapitulate why the scattering phase shift can be discontinuous as a result of Burgers circuit and then further use Burgers circuit to show its exact relation to density of states and partial density of states. Namely we can prove the reality of negative PDOS and prove the general regimes when semi classical FSR can be exact in a purely quantum regime. The advantage of using Burgers circuit is that these results can be shown to be very general and thus goes beyond all earlier works.

Besides we will show that the scattering phase shift behaviour of different experimentally \cite{schu,yang,yang1,kob1,kob2,hei,zaf} and theoretically \cite{deolegett,deo1,tan,eng,lee,levy,hack,singlechannel,3prong,solis,vass} studied mesoscopic systems can be understood from the Argand diagrams and analyticity. Each such system has its own peculiarities and so it is important to understand them with respect to a mathematical principle like Burgers circuit. Argand diagram for the scattering matrix element of these systems can be classified as, a) Argand diagram is closed, b) Argand diagram is open, c) Argand diagram encloses the phase singularity,
d) Argand diagram does not enclose the phase singularity, e) Argand diagram is simply connected and f) Argand diagram is multiply connected due to the presence of sub-loops. Specific properties of the scatterer only matter to the extent that the Argand diagram changes from one of these to another. Among them some changes are topologically possible and others are not. Understanding how an Argand diagram changes from one to the other explains most of the puzzles. 

In section \ref{model} we will analyse slips in the scattering phase shift \cite{schu,deolegett,tan,lee,levy}. 
In section \ref{inj} we will show that FSR too can be understood from Argand diagrams although its manifestation seems to be completely different for different potentials \cite{tan,lee,levy,singlechannel,3prong}. We will also show why FSR can become exact whenever there is a phase lapse. This is a physically counter-intuitive result that has been proven for particular potentials so far \cite{tan,lee,levy,singlechannel,satpathi}. We will show that this result depends on the properties of Argand diagram and hence very general and independent of the scattering potential. In section \ref{inj} we will use Burgers circuit to prove the possibility of negative partial density of states in real mesoscopic systems. To show negative PDOS by explicit calculation for any particular realistic potential is virtually impossible and so never shown before.    
\section{Burgers Circuit: an introduction}
Argand diagram is a plot of real versus imaginary parts of an analytic complex function and Burgers circuit is about phase changes and lapses being determined by phase singularities. 
Given a complex function $t$, if there is a phase singularity \cite{mic1} in its complex plane, then one can specify the strength of the singularity as follows \cite{mic1},
\begin{equation}
I=sgn Im\left( \nabla t^{*}\times\nabla t\right).\widehat{n} \label{one}
\end{equation}
$I$ is a topological quantum number which is always conserved in an interaction. $ I $ is a sign and so can be $ +1 $, $ -1 $ and $ 0 $. For a generalized `Burgers circuit' (BC) \cite{mic2},
\begin{equation}
\oint_{C} d\phi=2\pi I
\label{two}
\end{equation}
\begin{figure}
\centering
{\includegraphics[width=\textwidth ,keepaspectratio]{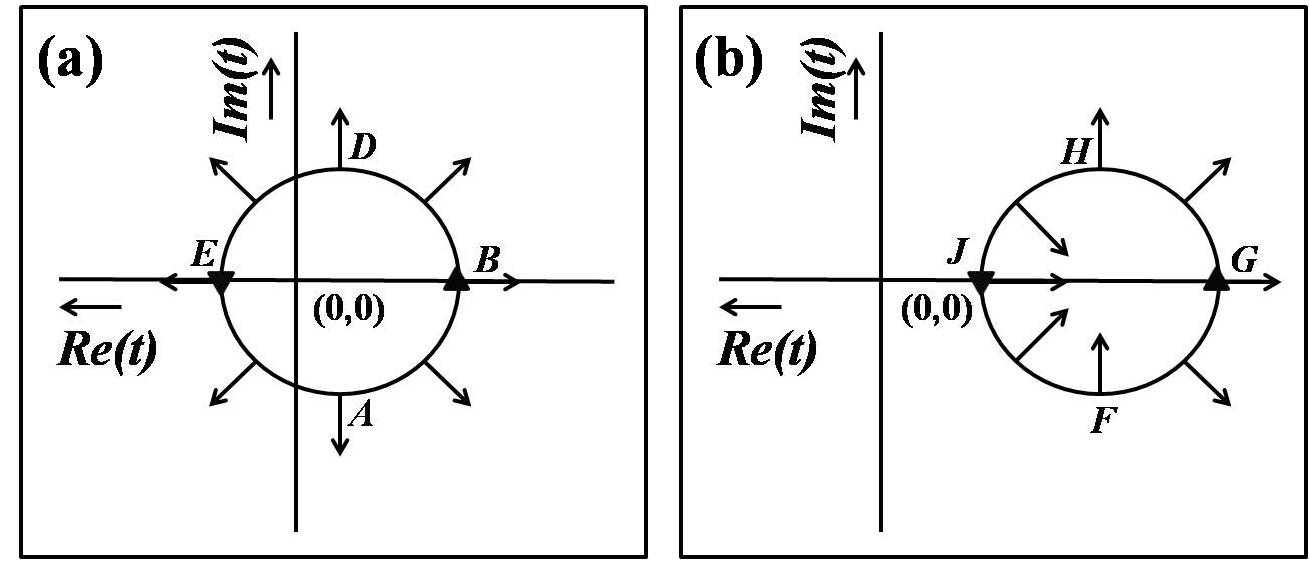}}
\caption{\label{f1}Schematic Argand diagrams that exemplify Eq. (\ref{two}). (a) When the Argand diagram contour encloses the singular point at $ (0,0) $ and (b) when the contour does not enclose the singular point.}
\end{figure}where, $ \phi=Arctan\frac{Im(t)}{Re(t)} $. If the contour $ C $ does not enclose the phase singularity then $I$ is 0. When the contour $ C $ enclosing a phase singularity is clockwise then $I$ is -1 and when the contour $ C $ enclosing a phase singularity is counter-clockwise then $I$ is +1. A scattering matrix element is a complex function for which Argand diagram can be drawn and concept of BC can be applied. For the rest of the paper where we refer to scattering matrix element, we will mean the scattering matrix element of a quantum mechanical particle say an electron. Argand diagram for a scattering matrix element is generally counter-clockwise, as incident energy of the scattering particle increases. At the phase singularity, $ Re(t)=0 $ and $ Im(t)=0 $, implying the phase singularity is at the origin. Figs. \ref{f1}(a) and \ref{f1}(b), shows schematic Argand diagrams for a complex function $ t $. Fig. \ref{f1}(a), shows a typical counter-clockwise contour (ABDEA) of an Argand diagram enclosing the phase singularity at the origin. Hence for this case, $ I = +1 $. The contour trajectory (ABDEA) is concave throughout with respect to the singular point at the origin. Following Eq. (\ref{two}), the net change in phase in tracing ABDEA, in Fig. \ref{f1}(a), is 2$\pi$. Thus the phase monotonously increases for a concave trajectory. Fig. \ref{f1}(b) shows a typical contour (FGHJF) of an Argand diagram not enclosing the phase singularity at the origin. Hence for this case, $ I =0 $. The contour has both concave (FGH) and convex (HJF) trajectories with respect to the singular point at the origin. Following Eq. (\ref{two}), the total change in phase in Fig. \ref{f1}(b) is zero. It is possible if the phase increases for the concave trajectory and decreases for the convex trajectory, the net phase change being zero. This can be easily verified by calculating $\phi$ at any point ($ Re(t), Im(t) $) of the trajectory and will be further demonstrated below. Thus for both Figs. \ref{f1}(a) and \ref{f1}(b), the phase change is determined by the singular point and the two follow the same principle. As a special case of the two, one can have a situation where the contour touches the singular point. In which case too the phase change can be understood from the same principle and will be explained later. Also if it is a closed contour as in Figs. \ref{f1}(a) and \ref{f1}(b), then Eq. (\ref{two}) is exact. In the real systems we will discuss in this work, the contours may not be closed. However, we will extend Eq. (\ref{two}) to understand such cases too.  

Whatever be the shape of a closed contour $ C $, the phase change is given by Eq. (\ref{two}). As a consequence, the real and imaginary parts of $t$ are not independent of each other but are related. Let us say the complex transmission amplitude be $ t(U)=\vert t(U)\vert e^{i\theta_{t}(U)} $ where $U$ can be any parameter like incident energy or gate voltage. Then
\begin{equation}
Re\left( t(U)\right)=\frac{1}{\pi} P \int_{-\infty}^\infty \frac{Im(t(U^{'}))}{U^{'}-U} dU^{'}			\label{three a}
\end{equation}
\begin{equation}
Im\left( t(U)\right)=-\frac{1}{\pi} P \int_{-\infty}^\infty \frac{Re(t(U^{'}))}{U^{'}-U} dU^{'}			\label{three b}
\end{equation}
These are the well known Kramers-Kronig relations \cite{jack}.
Another way in which the relation can be stated is in terms of Hilbert Transform \cite{eng},
\begin{equation}
ln\vert t(U)\vert =\frac{1}{\pi} P \int_{-\infty}^\infty \frac{\theta_{t}(U^{'})}{U^{'}-U} dU^{'}			\label{four a}
\end{equation}
\begin{equation}
\theta_{t}(U)=-\frac{1}{\pi} P \int_{-\infty}^\infty \frac{ln\vert t(U^{'})\vert}{U^{'}-U} dU^{'}			\label{four b}
\end{equation}
Since $ I $ is a conserved quantity, by adding terms to a Hamiltonian (or details to the states in the scatterer) we cannot remove the phase singularities of the wave function in the complex plane. Phase changes are determined by the phase singularities. Depending on the interaction the Argand diagram can however change and so a theoretical understanding of the experimental data may not crucially depend on the details of the sample or model. Sample details can change the shape of the contour $ C $, but as these theorems state, to understand the phase changes, we do not need all these details.
Englman and Yahalom \cite{eng} had shown that the experimental data for scattering phase shift and scattering cross section of a quantum dot, are consistent with Hilbert transforms. We will show that the principles of analyticity and Eq. (\ref{two}) can be used to arrive at our results.
\section{Model Potentials} \label{model}
In this section we analyse scattering phase shifts for different potentials that has been theoretically studied \cite{tan} so far, for analysing the experimental observations \cite{schu,yang,yang1}. We intend to analyse w.r.t Eq. (\ref{two}) which has not been done in earlier works. Again as explained with Eq. (\ref{two}), we do not need very complicated realistic potentials to understand the phase shifts but we need representative potentials that can be exactly solved and help us understand different aspects of Eq. (\ref{two}). 
\subsection{Double delta function potential in one dimension}\label{double}
\begin{figure}
\centering
{\includegraphics[width= \textwidth ,keepaspectratio]{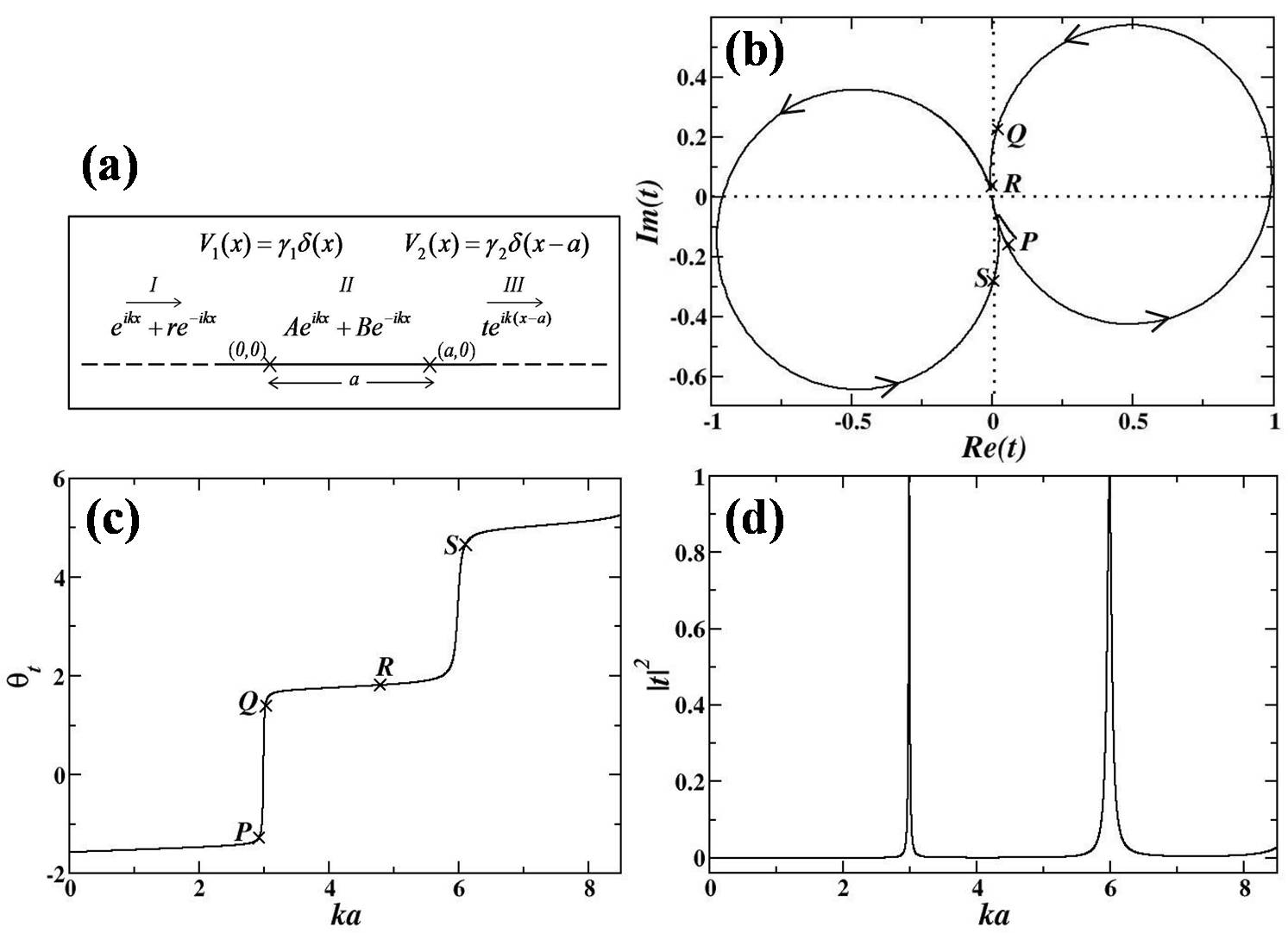}}
\caption{\label{f2}(a) Schematic representation for scattering of electrons by a double delta function potential in one dimension. The direction of incident and scattered electrons are represented by arrows. The solid line represents a quantum wire with double delta function potentials at positions $ x=0 $ and $ x=a $ respectively shown by cross (X) marks. $ \gamma_{1} $ and $ \gamma_{2} $ are the strengths of the potentials. The dashed lines represent the fact that the quantum wire is connected to electron reservoirs via leads. (b) Argand diagram for transmission amplitude for the double delta function potential. (c) Plot of transmission phase shift $ \theta_{t} $ versus $ ka $ and (d) plot of transmission coefficient $ |t|^2 $ versus $ ka $, for the double delta function potential using parameters $ e\gamma_{1}a=e\gamma_{2}a=40 $, $ a=1 $, $ e=1, 2m_{e}=1 $ and $ \hbar=1 $.}
\end{figure}

Let us first consider scattering by a double delta function potential in one dimension (1D) schematically shown in Fig. \ref{f2}(a), that was studied in ref. \cite{tan}. Although a simple potential, it exhibits pronounced Breit Wigner (BW) resonances. We will use Eq. (\ref{two}) to understand the scattering phase shift for this system and hence for BW resonances. The scattering potential for this system can be written as,
\begin{equation*}
V_{1}(x)=\gamma_{1}\delta(x)
\end{equation*}
\begin{equation*}
V_{2}(x)=\gamma_{2}\delta(x-a)
\end{equation*}
where $ \gamma_{1}$ and $\gamma_{2} $ are the strengths of potential $ V_{1}$ and $V_{2} $, respectively. The wave function in different regions marked $ I, II$ and $ III $ are (see Fig. \ref{f2}(a)),
\begin{equation*}
\psi(x)=\begin{cases}
e^{ikx}+r e^{-ikx}, \textit{for $x<0$,}\\
A e^{ikx}+B e^{-ikx}, \textit{for $0<x\leq a$,}\\
t e^{ik(x-a)}, \textit{for $x>a$.} \end{cases}
\end{equation*}
Here  $ r $ and $ t $ are the reflection and transmission amplitudes, $ k=\sqrt{\frac{2m_{e}}{\hbar^{2}} E}$ is the wave vector and $ E $ is incident Fermi energy. $ t=\mid t\mid e^{i\theta_{t}} $, where, $ \theta_{t} = Arctan\frac{Im(t)}{Re(t)} $ is transmission phase shift and $ |t|=\sqrt{Im(t)^2+Re(t)^2} $ is transmission modulus.
The Argand diagram for $ t $ is shown in Fig. \ref{f2}(b), where energy is varied to remain within the first Riemann surface. There is a phase singularity at the origin where $ t=0 $. The Argand diagram encloses the singularity but is not closed in the first Riemann surface. Figs. \ref{f2}(c) and \ref{f2}(d) shows the transmission phase shift $ \theta_{t} $ and the transmission coefficient $ |t|^2 $, respectively, as a function of $ ka $, using the same parameters as in Fig. \ref{f2}(b). 

In Fig. \ref{f2}(b), the contour starts from the origin where $ E=0 $, goes first through point $ P$ and then through $Q, R$ and $S$. The trajectory facing the singular point at the origin is concave throughout, and thus as discussed with Fig. \ref{f1}, the phase increases continuously. This is evident in Fig. \ref{f2}(c) where the points $P, Q, R$ and $S$ are also shown at their respective values of $ ka $. As the trajectory comes closer to the point of phase singularity, phase changes are very small and energy cost is very high. In Fig. \ref{f2}(b), the energy at the points marked $P, Q, R$ and $S$ are 8.614$ a^2 $, 9.12$ a^2 $, 23.61$ a^2 $ and 37.58$ a^2 $. Thus the energy change in going from $P$ to $Q$ (a large arc in the trajectory in Fig. \ref{f2}(b)) is very small, whereas the energy change in going from $Q$ to $R$ (a small arc in the trajectory in Fig. \ref{f2}(b)) is very high. The point $R$ is very close to the singular point. Thus it costs a lot of energy as the Argand diagram trajectory tries to approach the point of phase singularity. 
\subsection{Stub potential}\label{stub}
\begin{figure}
\centering
{\includegraphics[width=\textwidth ,keepaspectratio]{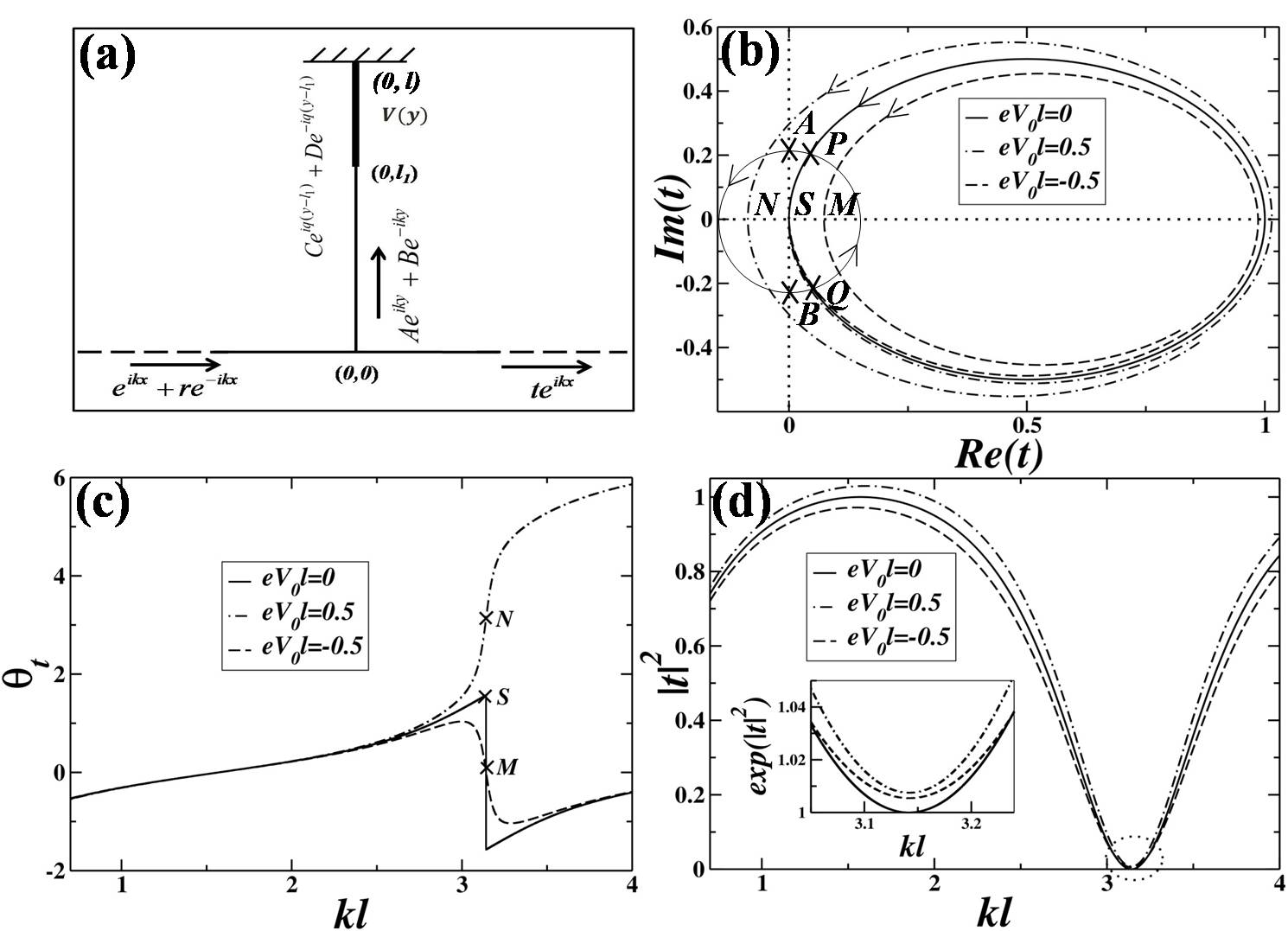}}
\caption{\label{f3}(a) Schematic representation for scattering of electrons by a stub potential. The direction of incident and scattered electrons are shown by arrows. The potential, represented by the bold line along y-axis, is $ V(y)=iV_0 $. (b) Argand diagram for transmission amplitude for different values of $ eV_0 l$, where, $ e $ is electronic charge and $ l $ is the length shown in (a). The thick solid line is Argand diagram for the case when $eV_{0}l=0$, the dot-dashed line is that for $eV_{0}l=0.5$ and the dashed line is for $eV_{0}l=-0.5$. Here $ l_{1}=.5l$, $l=1$, $ e=1, 2m_{e}=1 $ and $ \hbar=1 $. (c) Plot of transmission phase shift $ \theta_{t} $ and (d) plot of transmission coefficient $ \vert t \vert^{2} $, as a function of dimensionless wave vector $ kl $, taking the same notations and parameters as in Fig. \ref{f3}(b).}
\end{figure}

Another scattering potential often studied \cite{tan, deomplb} to understand the experiments \cite{schu,yang,yang1} is known as the stub, which is an infinite one-dimensional quantum wire with a finite side branch. A schematic representation of this system is shown in Fig. \ref{f3}(a). This system is topologically not the same as a one dimensional system, as in this case the origin $ (0,0) $ (see Fig. \ref{f3}(a)) is connected to three other directions. This potential shows discontinuous phase drops by $ \pi $ \cite{tan, deomplb} similar to that observed in experiments \cite{schu, yang1}. Prior to that this discontinuous scattering phase shift was shown to cause breakdown of parity effect of single particle states \cite{deolegett}. However, if there is an energy scale associated with such a sharp phase change has remained a puzzle \cite{schu}. Heiblum et al \cite{schu} quote that `The appearance of a second energy scale in the phase jump between resonances also cannot be understood...'. They also mention that it suggests existence of an unusually large energy scale.
We will use Eq. (\ref{two}) to understand why scattering phase shift can change discontinuously as some parameter is varied.  
Electrons are incident from left (see Fig. \ref{f3}(a)) with energy $ E $. The thin lines represent one dimensional quantum wires with zero potential, while the bold line represents quantum wire with a finite potential $ V(y) $ given by,
\begin{equation*}
V(y)= \begin{cases}0, \textit{for $0<y\leq l_{1}$,}\\
iV_{0}, \textit{for $l_{1}<y\leq l$,} \\
\infty, \textit{for $y>l$.} \end{cases}
\end{equation*}
The potential $ V(y) $ is taken to be imaginary, as it allows us to make the Argand diagram trajectory approach and cross the point of phase singularity. This cannot be done with real potentials as $ I $ in Eq. (\ref{two}) is a conserved quantity. Imaginary potentials are known as optical potentials \cite{jayan1, jayan2, jayan3, jayan4}. They are often used to simulate the effect of decoherence and open systems. If the system changes from a closed one to an open one then $ I $ may be different for the two cases. When $ V_{0} $ is 0 then $ V(y) $ is 0 and the system is a closed conserved system. And when $ V_{0} $ is non-zero then $ V(y)=iV_{0}$ and it is an open system. The wave function in the different regions are given by,
\begin{equation*}
\psi(x,y)=\begin{cases}
e^{ikx}+r e^{-ikx}, \textit{for $x<0$,}\\
t e^{ikx}, \textit{for $x>0$,}\\
A e^{iky}+B e^{-iky}, \textit{for $0<y\leq l_{1}$,}\\
C e^{iq(y-l_{1})}+D e^{-iq(y-l_{1})}, \textit{for $l_{1}<y<l$,} \\
0, \textit{for $y=l$.}\end{cases}
\end{equation*}
Here  $ r $ and $ t $ are the reflection and transmission amplitudes, $ k=\sqrt{\frac{2m_{e}}{\hbar^{2}} E} $ is the wave vector along thin lines, $ q=\sqrt{\frac{2m_{e}}{\hbar^{2}}(E-iV_{0})}$ is the wave vector along the bold line, and $ E $ is the Fermi energy.
Solving the scattering problem using Griffiths boundary conditions \cite{jayan1, jayan2, jayan3, jayan4}, that the wave function is continuous and the currents are conserved at the junction at $(0,0)$ in Fig. \ref{f3}(a), we get $ r $ and $ t $ as a function of energy, $ E $.
\begin{equation}
r=\frac{1-\frac{ik-q_{1}}{ik+q_{1}} e^{-2ikl_{1}}}{1+3 \frac{ik-q_{1}}{ik+q_{1}} e^{-2ikl_{1}}} \label{seven} 
\end{equation}
\begin{equation}
t=\frac{2(1+\frac{ik-q_{1}}{ik+q_{1}} e^{-2ikl_{1}})}{1+3 \frac{ik-q_{1}}{ik+q_{1}} e^{-2ikl_{1}}} \label{eight}
\end{equation}
where, $ q_{1}=q \left[ cot\left( {q(l-l_{1})}\right)\right] $. Transmission phase shift is given by $ \theta_{t}= Arctan\frac{Im(t)}{Re(t)} $. 
Fig. \ref{f3}(b) show the Argand diagrams for the transmission amplitudes, given by Eq. (\ref{eight}), for different values of $ eV_0l $, where energy is varied to remain within the first Riemann surface. The thick solid line is for $eV_{0}l=0$, the dot-dashed line is for $eV_{0}l=0.5$ and the dashed line is for $eV_{0}l=-0.5$. There is a phase singularity at $ t=0 $ marked as $S$ in Fig. \ref{f3}(b). Figs. \ref{f3}(c) and \ref{f3}(d) show the transmission phase shift $ \theta_{t} $ and transmission coefficient $ |t|^2 $, respectively, as a function of $ kl $ for different values of $ eV_{0}l $, using the same parameters and same notations as in Fig. \ref{f3}(b).

In Fig. \ref{f3}(b), if we draw a circle (thin solid line through $PABQP$) around the phase singularity, then Eq. (\ref{two}) implies the total phase change along the contour is $ 2\pi $. If the shape and size of the contour is altered, the net phase change along the contour remains the same provided the contour encircles the phase singularity. The phase change while going from $  Q$ to $  P$ along the arrow in the contour is less than $ \pi $ as the phase change in going from $  B$ to $A$ is $ \pi $. In the limit when the radius of the circle $ P, A, B, Q $ is tending to zero, $P$ approaches $A$, $Q$ approaches $B$ and $ P, A, B, Q $ all coincide with the singular point $ S $. In this limit going from $  Q$ to $  P$ would imply a discontinuous phase change of $ \pi $. Therefore a trajectory that tangentially touches the singular point $ S $ like the thick solid line for $ eV_0 l = 0 $ in Fig. \ref{f3}(b), will exhibit a discontinuous phase change of $ \pi $ at $ S $. This phase change can be seen in the solid line for $ eV_0 l = 0 $ in Fig. \ref{f3}(c) (at point $ S $ marked at the same value of $ kl $ as in Fig. \ref{f3}(b)), where we have plotted the transmission phase shift $ \theta_{t} $ versus $ kl $. 
We have already discussed in the previous section that as the Argand diagram trajectory approaches the point of phase singularity, the energy cost is very high. So, it is surprising that in a mesoscopic system, one can have such a high energy scale and one can realize discontinuous phase drops that are generally not seen for classical wave transport. In fact, it is not a real energy scale but an effective energy scale. Scattering by a stub of length $ l $ can be mapped into a problem of a delta function potential in one dimension where the strength of the delta function potential is $ k cot(kl) $. That is, $ V^{eff}(x)=kcot(kl)\delta(x) $ has same reflection amplitude $ r $ and transmission amplitude $ t $ as a stub of length $ l $. This is an effective potential that the electrons encounter while the real potential is $ V_0 = 0 $. At $ kl=\pi $, a small change in $ k $ or $ l $ means a very large change in the effective potential $ V^{eff}(x) $. Such large effective energy scales are known in other areas of condensed matter physics, for example, the effective electron mass becomes $ \infty $ at the band edge.

For $ eV_0 l = 0.5 $, the Argand diagram trajectory shown by the dot-dashed line in Fig. \ref{f3}(b), intersects the $ Re(t) $ axis at point $ N $ at $ kl=\pi $. The Argand diagram trajectory facing the phase singularity at $ S $ is concave throughout. So, Eq. (\ref{two}) implies a monotonously increasing phase. This phase behaviour can be seen in the dot-dashed line in Fig. \ref{f3}(c) where we have plotted the transmission phase shift $ \theta_{t} $ versus $ kl $ and marked $ N $ at $ kl=\pi$. For $eV_0 l= -0.5 $, the Argand diagram trajectory shown by the dashed line in Fig. \ref{f3}(b), intersects the $ Re(t) $ axis at point $ M $ at $ kl=\pi $. The Argand diagram trajectory facing the phase singularity is partially convex and partially concave. As discussed with respect to Eq. (\ref{two}), the phase increases for the concave part and decreases for the convex part of the trajectory. This phase behaviour can be seen in the dashed curve in Fig. \ref{f3}(c) where we have plotted the transmission phase shift $ \theta_{t} $ versus $ kl $ and marked the point $ M $ similarly. Therefore, if $ eV_0 l$ is continuously changed from $ 0.5 $ to $ -0.5 $, then the point $ N $ moves to $ M $ crossing the point of phase singularity. When $ eV_0 l$ becomes $ 0 $ we get the solid curves in Figs. \ref{f3}(b), \ref{f3}(c) and \ref{f3}(d) This implies that with an imaginary potential, just by changing sign of the potential it is possible to cross the singular point. $ I $ for dot-dashed line is 1 and that for the dashed line is 0. In other words by changing an imaginary potential we can make $ I $ change from 1 to 0. This is difficult with a real potential. If we tried to cross the singular point with real potentials it would cost an infinite amount of energy. We have thus discussed an effective real potential that can make an Argand diagram trajectory tangentially touch the point of phase singularity. Any effective potential that can make Argand diagram trajectory cross the point of phase singularity is not known.  

In Fig. \ref{f3}(c), we observe two different types of phase drops. For $ eV_{0}l=0 $, we get a discontinuous phase drop (shown by the solid curve in Fig. \ref{f3}(c)) and for $ eV_{0}l=-0.5 $, we get a gradual phase drop (shown by the dashed curve in Fig. \ref{f3}(c)). If we make $ eV_{0}l <-0.5 $, the point $ M $ in Fig. \ref{f3}(b) will shift more to the right and the phase drop will be less in magnitude and also less sharp. In Fig. \ref{f3}(d), using the same parameters and same notations as in Figs. \ref{f3}(b) and \ref{f3}(c), transmission coefficient $ |t|^{2} $ is plotted as a function of $ kl $ for different values of $ eV_{0}l $. The inset in Fig. \ref{f3}(d) shows the transmission coefficient $ |t|^{2} $ in exponential scale in the region around $ kl= \pi $ (shown by dotted circle). At $ kl=\pi $, the thick solid curve for $ eV_{0}l=0 $ goes to zero, while the dot-dashed and dashed curves for $ eV_{0}l=0.5 $ and $ eV_{0}l=-0.5 $ respectively, go through a non-zero minima and are very close. The phase behaviours for dot-dashed and dashed curves are completely different as can be seen from Fig. \ref{f3}(c). 

Thus in this section, we have explained using Argand diagram and Burgers circuit, why phase (scattering phase shift) drops can occur discontinuously? Why such a phase drop can disappear or change from discontinuous to gradual? The gradual phase drop in the dashed curve of Fig. \ref{f3}(c) at $ M $ is related to the analytic property of complex transmission amplitude and how the trajectory encloses the singularity. Whether the gradual drop is sharp or slow, depends on the distance $SM$ (Fig. \ref{f3}(b)) at which the convex trajectory intercepts the $Re(t)$ axis. 
\subsection{Single channel quantum wire}\label{singlechannel}
\begin{figure}
\centering
{\includegraphics[width=8cm,keepaspectratio]{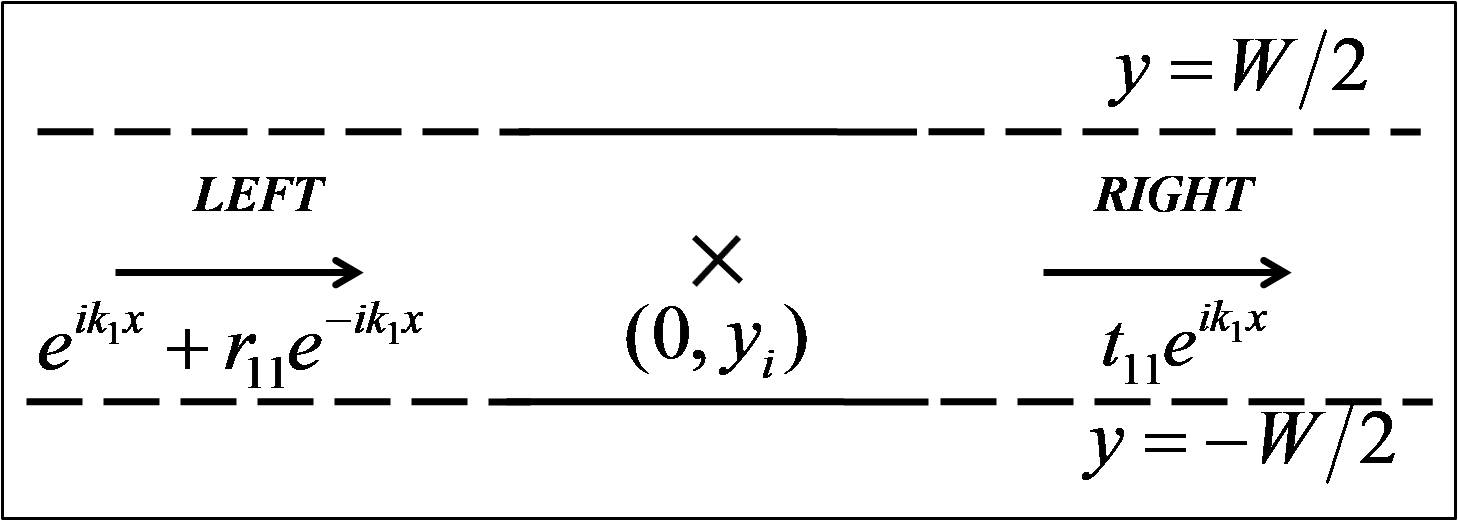}}
\caption{\label{singlechannelqw}Schematic representation for scattering of electrons by a one dimensional delta function potential. The position of a delta function potential is shown by cross (X) mark.}
\end{figure}
The next scattering potential we consider, is a delta function potential in a single channel quantum wire \cite{singlechannel,bagwell}. This essentially means, there is a single propagating channel while all other channels are evanescent. These evanescent channels are characteristic of quasi one dimension (Q1D), and make this scattering potential completely different from that of a delta function potential in one dimension \cite{gas1,gas2,gas3}. This system gained relevance with respect to the experiments of Schuster et al. \cite{schu}, Yang Ji et al. \cite{yang,yang1}, etc., because it too shows phase drops \cite{deo1} like those observed in case of the stub. The delta function potential can create a bound state in the continuum and it is this bound state that non-trivially affects the scattering. Any other potential in quasi one dimension that can sustain a bound state will produce similar effects. Essentially, such a bound state cause a Fano resonance which is at the heart of the features observed for this potential. In this section we will explain the scattering phase shift for this potential from Argand diagram. The system is shown in Fig. \ref{singlechannelqw}. The quantum wire is shown by solid line with a delta function potential at position (0, $y_i$) shown by a cross mark. $ W $ is width of the quantum wire. The dashed lines represent the fact that the quantum wire is connected to electron reservoirs via leads. Electrons are injected from the left reservoir into the left lead. The electrons are allowed to propagate along $ x $ direction, but confined along $ y $ direction. The confinement potential in the leads is taken to be hard wall potential and is given by,
\begin{figure}
\centering
{\includegraphics[width=\textwidth ,keepaspectratio]{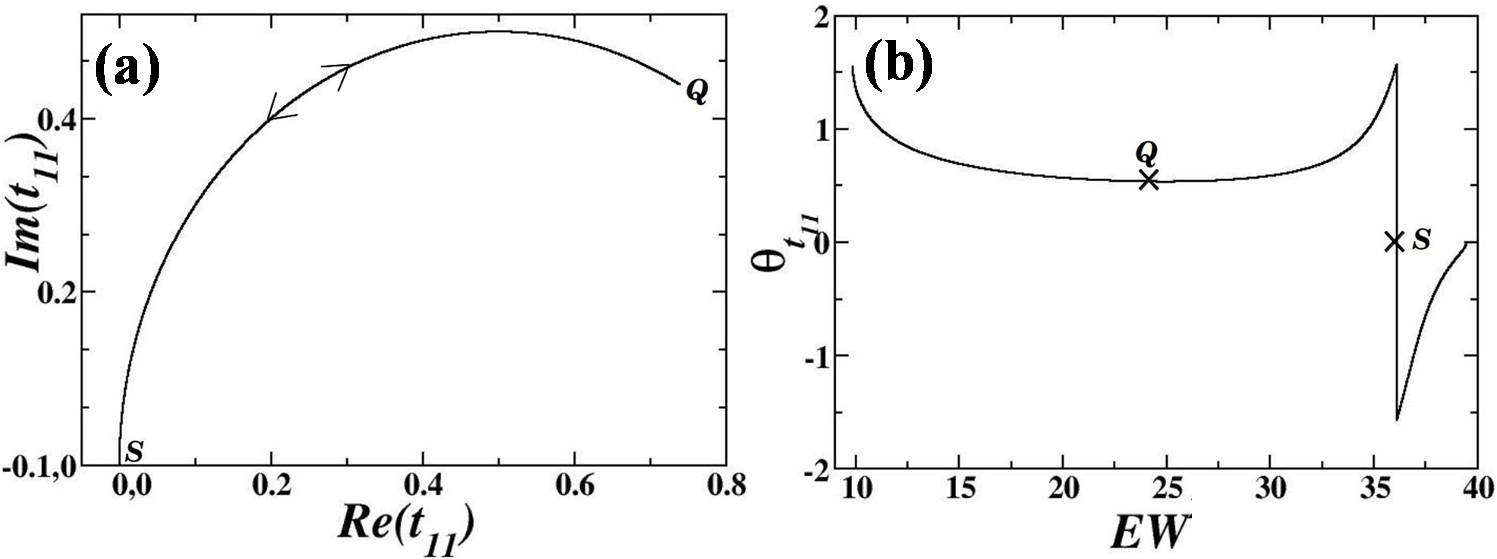}}
\caption{\label{f5}(a) Argand diagram for transmission amplitude and (b) plot of transmission phase shift $ \theta_{t_{11}} $ versus $EW$, for scattering by a delta function potential in a single channel quantum wire. Here $ e\gamma W = -1.5 $, $ y_i=0.21W, e=1, W=1 $ and we have considered 500 evanescent modes.}
\end{figure}
\begin{equation*}
V_c(y)=\begin{cases}
\infty, \textit{for $|y|\geq \frac{W}{2}$,}\\
0, \textit{for $|y|<\frac{W}{2}$.} \end{cases}
\end{equation*}
The direction of propagation is shown by arrows. 
The scattering potential shown by cross mark is given by,
\begin{equation*}
V(x,y)=\gamma\delta(x)\delta(y-y_i)
\end{equation*}
Here $ \gamma $ is the strength of the delta function potential. 
The asymptotic wave function in different regions are shown in Fig. \ref{singlechannelqw}. One can solve the scattering problem \cite{bagwell} to find,
\begin{equation}
r_{11}=-\frac{i\frac{\Gamma_{11}}{2k_{1}}}{1+\sum_{n\geq 2} \frac{\Gamma_{nn}}{2\kappa_n}+i\frac{\Gamma_{11}}{2k_{1}}} \label{r11}
\end{equation}
\begin{equation}
t_{11}=\frac{1+\sum_{n\geq 2} \frac{\Gamma_{nn}}{2\kappa_n}}{1+\sum_{n\geq 2} \frac{\Gamma_{nn}}{2\kappa_e}+i\frac{\Gamma_{11}}{2k_{1}}} \label{t11}
\end{equation}
Here  $ r_{11} $ and $ t_{11} $ are the reflection and transmission amplitudes and $\Gamma_{nm}$ is given by,
\begin{equation}
\Gamma_{nm}=\gamma sin\left[\frac{n\pi}{W}\left(y_{i}+\frac{W}{2} \right)  \right]sin\left[\frac{m\pi}{W}\left(y_{i}+\frac{W}{2} \right)\right] \label{gnn}
\end{equation}
where, $ m $ and $ n $ are integers. $ k_1=\sqrt{\frac{2m_{e}}{\hbar^{2}}E-\frac{\pi^{2}}{W^{2}}}$ is the wave vector for the propagating channel, $ \sum_n $ denotes sum over evanescent modes, $ \kappa_{n}=\sqrt{\frac{n^{2}\pi^{2}}{W^{2}}-\frac{2m_{e}}{\hbar^{2}}E} $ where $ n $ takes values $ 2, 3, ... \infty $ and $ E $ is the incident Fermi energy. Transmission phase shift is given by, $ \theta_{t_{11}}=Arctan\frac{Im(t_{11})}{Re(t_{11})} $. 

The Argand diagram for transmission amplitude $ t_{11} $ is shown in Fig. \ref{f5}(a). There is a phase singularity at the origin where, $ t_{11}= 0 $ (shown by the point $ S $). Fig. \ref{f5}(b) shows  transmission phase shift $ \theta_{t_{11}} $ as a function of energy, for the same parameters as in Fig. \ref{f5}(a). In Fig. \ref{f5}(a), the trajectory (shown by thick solid line) starts from the origin (point marked $ S $) goes upto point $ Q $, and traces back the same path to pass the origin making $SQS$ a closed contour. The direction of the trajectory is therefore shown by a double headed arrow. At the energy, where the trajectory goes from $ Q $ to $ S $ and touches the point of phase singularity at origin, i.e. point $ S $, we expect a discontinuous phase drop of $ \pi $, following the same argument as in the case of the stub. This phase drop can be seen in Fig. \ref{f5}(b), where the points $ Q$ and $ S $ are also marked at their respective energies. Thus the discontinuous phase drop is a natural consequence of Eq. (\ref{two}).
\subsection{Three prong potential}\label{threeprong}
\begin{figure}
\centering
{\includegraphics[width=.45\textwidth ,keepaspectratio]{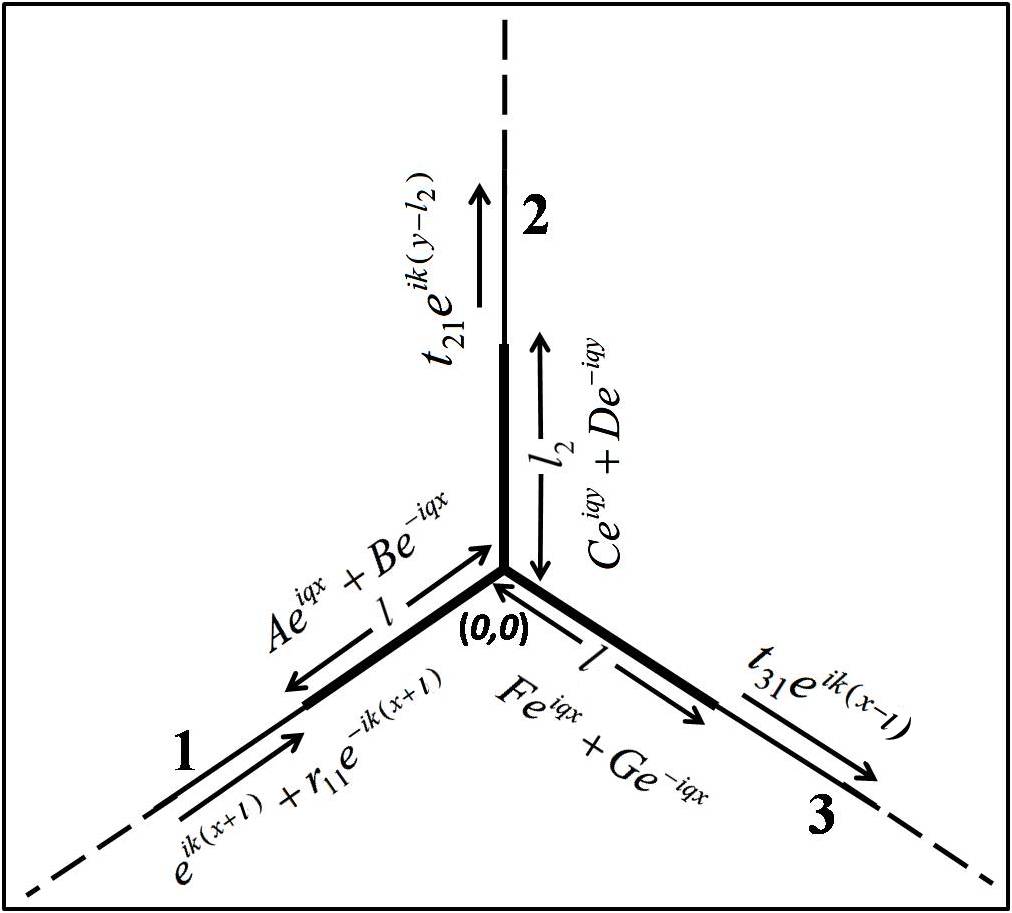}}
\caption{\label{fig 3pr}Schematic representation of scattering of electrons by a three prong potential. The direction of incident and scattered electrons are shown by arrow heads. The potential is non-zero along the bold lines of lengths $ l $, $ l_2 $ and $ l $ along $ -x, +y $ and $ +x $ axes, respectively. }
\end{figure}

We now consider another potential called the three prong potential \cite{3prong}. This potential will help us to demonstrate other non-trivial aspects that follow from Eq. (\ref{two}). A schematic representation of the three prong potential is shown in Fig. \ref{fig 3pr}.
The thin lines represent one dimensional quantum wires with potential $ V=0 $, and the bold lines represent quantum wires with non zero potential, i.e., $ V\neq0 $. The arms of the prong are labelled as 1, 2 and 3 as shown in Fig. \ref{fig 3pr}. The electrons are considered to be incident from left, the direction of incidence being shown by arrows. The wave function in the different regions are given by,
\begin{figure}[h!]
\centering
{\includegraphics[width=\textwidth, keepaspectratio]{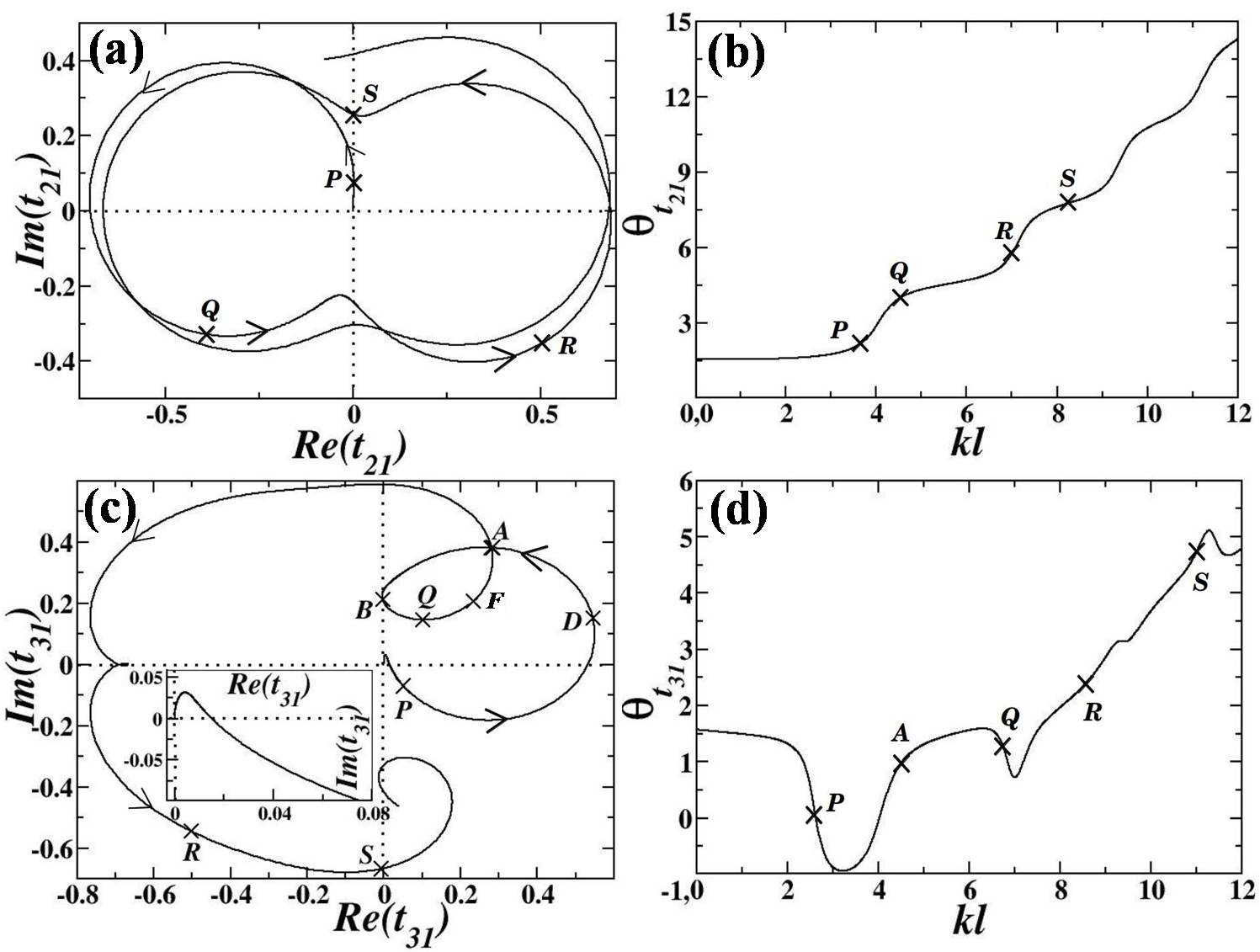}}
\caption{\label{f7}(a) Argand diagram for transmission amplitude $ t_{21} $ and (b) plot of scattering phase shift $\theta_{t_{21}} $ as a function of $ kl $ varying the wave vector from $ 0 $ to $ 12 $. (c)  Argand diagram for transmission amplitude $ t_{31} $  and (d) plot of scattering phase shift $ \theta_{t_{31}} $ as a function of $ kl $ varying the wave vector from $ 0 $ to $ 12 $. For all the figures, $ l=1$, $ l_{2}=5 l$, $ e=1 $ and $ eVl= -1000 $.} 
\end{figure}
\begin{equation*}
\psi(x,y,z)=\begin{cases}
e^{ik(x+l)}+r_{11} e^{-ik(x+l)}, \textit{for $x<-l$,}\\
A e^{iqx}+B e^{-iqx}, \textit{for $-l<x<0$,}\\
C e^{iqy}+D e^{-iqy}, \textit{for $0<y<l_2$,}\\
F e^{iqx}+G e^{-iqx}, \textit{for $0<x<l$,}\\
t_{21} e^{ik(y-l_2)}, \textit{for $y>l_2$,}\\
t_{31} e^{ik(x-l)}, \textit{for $x>l$.}\end{cases}
\end{equation*}
where $ k=\sqrt{\frac{2m_{e}}{\hbar^{2}} E}$ is the wave vector along the thin lines, $ q=\sqrt{\frac{2m_{e}}{\hbar^{2}} (E-V)}$ is the wave vector along the bold lines and $ E $ is the Fermi energy. Here $ r_{11} $ is the reflection amplitude for electrons incident from 1 and reflected back to 1, $ t_{21} $ is the transmission amplitude for electrons incident from 1 and transmitted to 2 and $ t_{31} $ is the transmission amplitude for electrons incident from 1 and transmitted to 3. These scattering matrix elements can be solved using Griffiths boundary conditions \cite{jayan1, jayan2, jayan3, jayan4}. The respective transmission phase shifts are given by, $\theta_{r_{11}}=Arctan\frac{Im(r_{11})}{Re(r_{11})} $, $ \theta_{t_{21}}=Arctan\frac{Im(t_{21})}{Re(t_{21})} $ and $ \theta_{t_{31}}=Arctan\frac{Im(t_{31})}{Re(t_{31})} $. 

Fig. \ref{f7}(a) shows the Argand diagram for transmission amplitude $ t_{21} $. There is a phase singularity at the origin, where $t_{21}=0$. In Fig. \ref{f7}(a) the trajectory of Argand diagram for $ t_{21} $ starts from the origin, goes through $P$ and then through $Q, R$ and $S$ following counter-clockwise direction shown by arrows. The trajectory is concave throughout and Eq. (\ref{two}) implies monotonously increasing phase. This monotonously increasing phase can be seen in Fig. \ref{f7}(b), where scattering phase shift $ \theta_{t_{21}} $ is plotted as a function of $ kl $ and here also the points $P, Q, R$ and $S$ are marked at the corresponding values of $ kl $. 

The Argand diagram for $ t_{31} $ shows something interesting. This Argand diagram is shown in Fig. \ref{f7}(c). There is a phase singularity at the origin, where $t_{31}=0$. In Fig. \ref{f7}(c), the trajectory of Argand diagram for $ t_{31} $ starts from origin, goes through $P$ and then through $D, A, B, Q, F, R$ and $S$, following counter-clockwise direction shown by arrows. Here interestingly, the trajectory develops a sub-loop $ABQFA$. This sub-loop results in a convex arc $ BQF $ in the trajectory that does not go through the origin. As explained earlier, there will be a gradual phase drop whenever such a convex arc is observed, following Eq. (\ref{two}). This can be seen in Fig. \ref{f7}(d), where scattering phase shift $ \theta_{t_{31}} $ is plotted as a function of $ kl$ and here also the points $P, A, Q, R$ and $S$ are marked. Presence or absence of such a sub-loop has no consequence on the line integral of phase along $ PDABQFRS $. This is because the contribution to the line integral coming from the sub-loop $ ABQFA $ is 0 and its presence or absence has no bearing on the value of $ I $. 
So, such a sub-loop as $ ABQFA $ in Fig. \ref{f7}(c) can appear or disappear as some parameter is varied as will be demonstrated in the next section. 

\section{Injectance and Friedel sum rule} \label{inj}
Local partial density of states (LPDOS) is defined as \cite{buttiker2,buttiker3,buttiker4,buttiker1}
\begin{equation}
\rho '(\alpha, \textbf{r}, \beta)= -\frac{1}{4\pi i}\left( s_{\alpha\beta}^{\dagger} \frac{\delta s_{\alpha\beta}}{\delta V(\textbf{r})} - s_{\alpha\beta} \frac{\delta s_{\alpha\beta}^{\dagger}}{\delta V(\textbf{r})}\right) 
\label{buttiker1}
\end{equation}
Here, $ s_{\alpha\beta}=\mid s_{\alpha\beta}\mid e^{i\theta_{s_{\alpha\beta}}}$ is the scattering matrix element for electrons incident from channel $ \beta $ and transmitted to channel $ \alpha $ and $ \frac{\delta}{\delta V(\textbf{r})} $ stands for a functional derivative with respect to the local potential $ V(\textbf{r}) $. Time spent at $ \textbf{r} $ by an electron going from channel $ \beta $ to $ \alpha $ is given by \cite{buttiker1}
\begin{equation*}
\tau '(\alpha, \textbf{r}, \beta)=\frac{h}{|s_{\alpha\beta}|^{2}}\rho '(\alpha, \textbf{r}, \beta) 
\end{equation*}
Therefore, time spent by an electron going from channel $ \beta $ to $ \alpha $, within the scattering region is given by \cite{buttiker1}
\begin{equation}
\tau(\alpha,\beta)=\frac{h}{|s_{\alpha\beta}|^{2}}\int_{sc reg} {d\textbf{r}} ^{3} \rho '(\alpha, \textbf{r}, \beta) \label{time1} 
\end{equation}
where, `screg' stands for scattering region. For a mesoscopic sample coupled to leads, this scattering region is by definition the sample \cite{buttiker1}. So, partial density of states (PDOS) of a mesoscopic sample is defined as \cite{buttiker1}
\begin{eqnarray}
\rho(\alpha, \beta)&=& -\frac{1}{4\pi i}\int_{sample} d\textbf{r}^{3} {\left( s_{\alpha\beta}^{\dagger} \frac{\delta s_{\alpha\beta}}{\delta V(\textbf{r})} - s_{\alpha\beta} \frac{\delta s_{\alpha\beta}^{\dagger}}{\delta V(\textbf{r})}\right) }\label{buttiker2}\\
\text{or,}\hspace{.2cm}  \rho(\alpha, \beta)&=& -\frac{1}{2\pi}\int_{sample} d\textbf{r}^{3}{\left( |s_{\alpha\beta}|^{2} \frac{\delta \theta_{s_{\alpha\beta}}}{\delta V(\textbf{r})}\right)}\label{buttiker2a}
\end{eqnarray}
PDOS are quite physical and manifests in a variety of experimental situations in mesoscopic systems \cite{buttiker3,buttiker4,buttiker1}. For example, decoherence in the scattering region is proportional to the time electrons spend in the scattering region. As another example, consider a sinusoidal voltage of frequency $ \omega $, $ V_{\beta}(\omega) $ applied at incident lead $ \beta $. The current measured at lead $ \alpha $ will be \cite{buttiker1},
\begin{eqnarray}
I_{\alpha}(\omega)=G_{\alpha\beta}(\omega)V_{\beta}(\omega) \label{buttiker9}
\end{eqnarray}
where, $ G_{\alpha\beta}(\omega) $ is the dynamical conductance matrix and is given by \cite{buttiker1},
\begin{eqnarray}
G_{\alpha\beta}(\omega)=G_{\alpha\beta}^{0}-i\omega E_{\alpha\beta}+K_{\alpha\beta}\omega^{2}+O(\omega^{3}) \label{buttiker10}
\end{eqnarray}
$ G_{\alpha\beta}^{0} $ is the dc-conductance matrix. $ E_{\alpha\beta} $ is proportional to $ \omega $ and governs the displacement currents and is given by \cite{buttiker1},
\begin{eqnarray}
E_{\alpha\beta}=e^{2}\rho(\alpha, \beta)-e^{2} \int d\textbf{r}' \rho(\alpha, \textbf{r}')\int d\textbf{r} g(\textbf{r},\textbf{r}')\rho(\textbf{r},\beta)\label{buttiker11}
\end{eqnarray}
where, $ g(\textbf{r},\textbf{r}') $ is the effective interaction potential. All these experimental situations explicitly involve $ \beta $ as the incoming channel and $ \alpha $ as the outgoing channel. All analysis of $ \rho(\alpha, \beta) $ will be made w.r.t the R.H.S of Eqs. (\ref{buttiker2}) and (\ref{buttiker2a}). $ \rho(\alpha, \beta) $ is theoretically undefined when the experiment does not clearly involve an incoming channel and an outgoing channel. This has to do with the fact that quantum mechanics is necessary (to explain experimental observations) but not sufficient. There are many alternate approaches that give similar results as quantum mechanics and merits and demerits of such alternate approaches is a never ending discussion that we will avoid in this work. The fact remains that results deduced from quantum mechanics has never been violated in an experiment. So, whatever we prove for the mathematical expression on the R.H.S of Eqs. (\ref{buttiker2}) and (\ref{buttiker2a}) will have consequences on experimental observations.

The integration over $\textbf{r}$ in Eq. (\ref{buttiker2}) can easily be done for a global change (for all $ \textbf{r} $ in the sample as well as in the leads) in $V(\textbf{r})$ by a constant amount $ \epsilon $, i.e. $\delta V(\textbf{r})=\epsilon$ for all $ \textbf{r} $. Such a constant global increase in potential is equivalent to decrease in incident energy $ E $, i.e.,
\begin{eqnarray}
\int_{global} d\textbf{r}^{3}\frac{\delta}{\delta V(\textbf{r})}\equiv -\frac{d}{dE} \label{buttiker15}
\end{eqnarray}
and, therefore,
\begin{eqnarray}
\int_{sample} d\textbf{r}^{3}\frac{\delta}{\delta V(\textbf{r})} \cong -\frac{d}{dE}\label{buttiker16}
\end{eqnarray}
is expected to work in the semi-classical limit \cite{buttiker2}. So, from Eq. (\ref{buttiker2})
\begin{eqnarray}
\rho(\alpha, \beta)&=&-\frac{1}{4\pi i}\int_{sample} d\textbf{r}^{3} {\left( s_{\alpha\beta}^{\dagger} \frac{\delta s_{\alpha\beta}}{\delta V(\textbf{r})} - s_{\alpha\beta} \frac{\delta s_{\alpha\beta}^{\dagger}}{\delta V(\textbf{r})}\right) }\nonumber\\
&\approx & -\frac{1}{4\pi i}\int_{global} d\textbf{r}^{3} {\left( s_{\alpha\beta}^{\dagger} \frac{\delta s_{\alpha\beta}}{\delta V(\textbf{r})} - s_{\alpha\beta} \frac{\delta s_{\alpha\beta}^{\dagger}}{\delta V(\textbf{r})}\right) }\nonumber\\
\text{or,}\hspace{.2cm} \rho(\alpha, \beta)&\approx & \frac{1}{4\pi i}{\left( s_{\alpha\beta}^{\dagger} \frac{d s_{\alpha\beta}}{dE} - s_{\alpha\beta} \frac{d s_{\alpha\beta}^{\dagger}}{dE}\right) } \hspace{1cm} (\text{Using Eq. (\ref{buttiker15})}) \nonumber
\end{eqnarray}
On simplifying we get,
\begin{eqnarray}
\rho(\alpha, \beta)&\approx & \frac{1}{2\pi}{\left( |s_{\alpha\beta}|^{2} \frac{d \theta_{s_{\alpha\beta}}}{dE}\right)}
\label{buttiker3}
\end{eqnarray}

It is known that $ \frac{d \theta_{s_{\alpha\beta}}}{dE} $ can be negative. Concluding $ \rho(\alpha, \beta) $ to be negative when R.H.S of Eq. (\ref{buttiker3}) is negative is completely wrong. Eq. (\ref{buttiker3}) is an approximate equality, which implies if R.H.S is negative, the L.H.S is not necessarily negative. Although PDOS $ \rho(\alpha, \beta) $ can be in principle negative as can be demonstrated from R.H.S of Eq. (\ref{buttiker2}) for some strictly 1D simple potentials (like a square well and square barrier) in very low energy regime. 1D potentials are an idealization and not physical and neither one can go to the necessary low energy regime in an experiment. This work is based on the realization that some general conclusions can be drawn about the R.H.S of Eqs. (\ref{buttiker2}) and (\ref{buttiker2a}) using the properties of Argand diagram and Burgers circuit which is true for any potential in 1D or in Q1D.

The problem of negative PDOS was theoretically studied in two different ways strictly in 1D. One is negative PDOS and the other is negative time scales. Measured time scales in real systems did not lead to any unique physical understanding and a review on the topic can be seen in ref. \cite{landauer1}. Ref. \cite{landauer1} did conclude that $ \rho(\alpha, \beta) $ or $ \tau(\alpha, \beta) $ are physical and later on found to be so \cite{buttiker1}. But regimes where they become negative has not received any theoretical attention beyond 1D. In other words negative $ \rho(\alpha, \beta) $ has not yet been shown for any physical system. 

Although PDOS as defined in Eq. (\ref{buttiker2}) can be negative, they add up to give the correct DOS \cite{buttiker1} which is positive. One can sum the PDOS (given by Eq. \ref{buttiker2}) over $ \alpha$ to get injectance \cite{buttiker2,buttiker1,satpathi} of lead $ \beta $,
\begin{eqnarray}
\rho(\beta)=-\frac{1}{4\pi i}\sum_{\alpha} \int_{sample} d\textbf{r}^{3} {\left( s_{\alpha\beta}^{\dagger} \frac{\delta s_{\alpha\beta}}{\delta V(\textbf{r})} - s_{\alpha\beta} \frac{\delta s_{\alpha\beta}^{\dagger}}{\delta V(\textbf{r})}\right) }\label{buttiker4a}
\end{eqnarray}
This is a measure of current delivered by the electrons incident along lead $ \beta $ and outgoing through all the leads. Similarly injectance can be defined for all possible leads and they are completely independent of each other. For example, the scattering problem depicted in Fig. \ref{fig 3pr} can define $ \rho(1) $. To get $ \rho(2) $ one has to solve a completely different scattering problem where the incident particle is from lead 2.
Using Eq. (\ref{buttiker16}) we get the semi-classical limit of Eq. (\ref{buttiker4a}),
\begin{eqnarray}
\rho(\beta)\approx \frac{1}{4\pi i}\sum_{\alpha} {\left( s_{\alpha\beta}^{\dagger} \frac{d s_{\alpha\beta}}{dE} - s_{\alpha\beta} \frac{d s_{\alpha\beta}^{\dagger}}{dE}\right) }\label{buttiker6a}
\end{eqnarray}
Summing $ \rho(\beta) $ over the $ \beta $ independent channels, we can obtain density of states (DOS) $ \rho(E) $, i.e., from Eq. (\ref{buttiker4a}),
\begin{equation}
\rho(E)= -\frac{1}{4\pi i}\sum_{\alpha\beta} \int_{sample} d\textbf{r}^{3} {\left( s_{\alpha\beta}^{\dagger} \frac{\delta s_{\alpha\beta}}{\delta V(\textbf{r})} - s_{\alpha\beta} \frac{\delta s_{\alpha\beta}^{\dagger}}{\delta V(\textbf{r})}\right) } \label{buttiker4}
\end{equation} 
and in the semi-classical limit given by Eq. (\ref{buttiker16}), we get,
\begin{equation}
\rho(E)\approx \frac{1}{4\pi i}\sum_{\alpha\beta} {\left( s_{\alpha\beta}^{\dagger} \frac{d s_{\alpha\beta}}{dE} - s_{\alpha\beta} \frac{d s_{\alpha\beta}^{\dagger}}{dE}\right) } \label{buttiker6}
\end{equation} 
Further simplification of R.H.S in Eq. (\ref{buttiker6}) gives
\begin{eqnarray}
\pi\rho(E)\approx\frac{d}{dE}\theta_f(E)
\label{friedel}
\end{eqnarray}
This is Friedel sum rule (FSR), where $ \theta_f(E)={1\over 2i}\frac{d}{dE}log (det[S]) $ is the Friedel phase, $S$ is the scattering matrix and $ \rho(E)=\frac{dN(E)}{dE} $ is density of states. Since injectance of all leads are independent while they add up to give DOS, it is important to understand injectance in order to understand FSR. So we will restrict our study to injectance.

The potentials in sections \ref{stub}, \ref{singlechannel} and \ref{threeprong} are typical examples of mesoscopic systems, and as reported in ref. \cite{tan,singlechannel,3prong}, FSR (or injectance) manifests in these systems, in different ways. Semi-classical regime being expressed by Eq. (\ref{buttiker16}) does not seem to be sufficient. Sometimes, FSR (or injectance) is exact at all energies (for example the stub \cite{tan}) and sometimes it is exact at an energy where quantum fluctuations dominate \cite{singlechannel}. There is a huge amount of system to system variation. 
It has been proved very generally that when the phase drops of $ \pi $ are discontinuous like that in the solid curve in Fig. \ref{f3}(c), then at the energy corresponding to this drop, semi-classical injectance will become exact \cite{singlechannel}. But when the phase drops are gradual like the dashed curve in Fig. \ref{f3}(c), this has not been proved in general but shown for particular cases \cite{singlechannel,satpathi}. Burgers circuit will help us to derive in this work the general connection between gradual phase drops and exactness of semi-classical injectance for any arbitrary potential that can exhibit such a gradual phase drop.

We will first prove that in realistic mesoscopic systems one can definitely observe negative PDOS. We will show that the R.H.S of Eqs. (\ref{buttiker2}) and (\ref{buttiker2a}) can become negative for a realistic mesoscopic system. We will also show that such a conclusion cannot be drawn in 1D that has been extensively studied before \cite{landauer1}. Also we will show that when there are such negative slopes in scattering phase shift of mesoscopic systems then semi-classical FSR can become exact in a quantum regime. Once again our proofs will depend on Argand diagram and Eq. (\ref{two}) and so our proof will be general and not depend on the specific properties of the scattering potential.

For this we first make the connection between Eq. (\ref{two}) and injectance. When $ S $ matrix is $ 2\times2 $ (for example, the cases of double delta function potential (Fig. \ref{f2}(a)), stub (Fig. \ref{f3}(a)), single channel quantum wire (Fig. \ref{singlechannelqw})), Eq. (\ref{friedel}) simplifies to,
\begin{eqnarray}
\frac{d}{dE}\theta_t(E)\approx\pi\rho(E) \label{self}
\end{eqnarray}
Here $t$ is the transmission amplitude and $ \theta_t=Arctan\frac{Im(t)}{Re(t)} $. It has been proved that this equation is valid even if $ \theta_{t} $ is discontinuous with $ E $ because $ \theta_{t} $ is analytic \cite{n1} wherein the R.H.S derivative of $ \theta_{t} $ is the same as the L.H.S derivative at the discontinuity.
Suppose when the energy is varied from $ 0 $ to $ E_1 $, then the Argand diagram for a typical scattering matrix element $ t $ traces a closed contour $C$. For such a case one can state that, Eq. (\ref{two}) takes the form
\begin{equation}
\oint_{C} d\theta_t=2\pi N(E_1) \label{nine}
\end{equation}
where, $ N(E_1) $ is number of states (obtained by integrating DOS $ \rho(E) $ from $ 0 $ to $ E_1 $) below energy $ E_1 $ and is to be identified with the conserved quantity $ I $. Comparing with Eq. (\ref{two}) we see $ \phi\equiv\theta_t $  and $ I\equiv{N(E_1)}$.  
If $C$ happens to be a completely closed contour, Eq. (\ref{nine}) is exact as it is equivalent to Eq. (\ref{two}). Any complex function or its phase has to satisfy Eq. (\ref{two}) and a scattering matrix element is no exception provided it is analytic. This analyticity is the basic requirement for Eq. (\ref{buttiker2a}) (PDOS) or Eq. (\ref{buttiker4a}) (injectance). However, in most cases of scattering problems, $C$ is not completely closed and when $ C $ is not completely closed one cannot expect any conserved quantity. In Fig. \ref{f2}(b), there is a phase singularity at the origin and the contour enclosing the singularity is not closed in the first Riemann surface. One can restrict the discussion to the first Riemann surface to understand the injectance. When the contour continue to the second Riemann surface, then the contour integral starts including contribution from the second phase singularity in the second Riemann surface. And then one has to extend the discussions here to include the effect of the second singularity. This does not change the arguments given here except that sometimes the error from the first Riemann surface can cancel the error from the second Riemann surface which need not be a systematic behaviour. The contour $C$ in Fig. \ref{f2}(b) starts from origin with zero energy and ends in the first Riemann surface at point marked as $S$, where the energy is $ E_1 $ (say). It is now known that (see Eq. (6) in ref. \cite{levy}),
\begin{eqnarray}
\int_{C'} d\theta_t &=& \int_0 ^ {E_1} \left[ \pi {dN(E)\over dE}- ImTr\left( G^a\frac{\partial {F}^a}{\partial E}\right)  \right] dE \label{self1}
\end{eqnarray}
We will replace $ C $ by $ C' $, when the contour is not completely closed. $ G^a $ is the advanced Greens function and $ {F}^a $ is self energy due to coupling the system with the leads. One can then state that $\int_0 ^ {E_1}ImTr\left( G^a\frac{\partial {F}^a}{\partial E}\right)dE $ is the correction term for Eq. (\ref{nine}) when contour $ C' $ is not closed. This statement can be alternately justified as follows. When the self energy is independent of incident energy, then the contour $ C $ is closed as well as the correction term is zero implying Eq. (\ref{self1}) becomes Eq. (\ref{nine}). For a double delta function potential in one dimension the correction term is very important to consider. There, energy dependence of self energy can be seen very easily in the broadening of consecutive resonance peaks (shown in Fig. \ref{f2}(d)). Therefore, one can refine the statement in Eq. (\ref{buttiker16}) to state,
\begin{eqnarray}
 \oint_{C}dE\int_{sample} dr^3 \frac{\delta}{\delta V(r)}=\oint_{C}dE \left( -\frac{d}{dE}\right) \label{buttiker17}
\end{eqnarray} 
where, we generate a closed contour $ C $ in the Argand diagram by varying energy $ E $ from $ 0 $ to $ E_1 $. Without this equality one cannot get Eq. (\ref{nine}) as can be easily verified. Now from Eq. (\ref{buttiker15}) one can state,
\begin{eqnarray}
\oint_{C}dE\int_{sample} dr^3 \frac{\delta}{\delta V(r)}=\oint_{C}dE\int_{global} dr^3 \frac{\delta}{\delta V(r)} \label{buttiker18}
\end{eqnarray}

In case of the solid line in Fig. \ref{f3}(b), the potential everywhere is real and is $ 0 $. It traces a closed contour in the first Riemann surface. Thus for this system Eq. (\ref{nine}) will become applicable. Explicit calculations of density of states \cite{tan} for the stub show this and so everything is consistent with Eq. (\ref{two}). However, for scattering by a delta function potential (Fig. \ref{singlechannelqw}) in a single channel quantum wire we get a counter intuitive result. In this case, the contour of the Argand diagram (shown in Fig. \ref{f5}(a)) is closed in a special way and so we expect Eq. (\ref{nine}) to be applicable. But explicit calculations of density of states \cite{singlechannel} show, that is not the case. It has been shown earlier that for this system the fundamental theorem of B\"{u}ttiker-Thomas-Pretre (BTP) also breaks down \cite{deo} due to the non-analyticity of scattering matrix elements. This is a consequence of the fact that delta function potential in Q1D incorporates a log divergence in scattering matrix elements. Hence this is a situation where Eq. (\ref{two}) cannot be applied.

For the three prong potential, shown in Fig. \ref{fig 3pr}, the scattering matrix is $ 3\times3 $ and the correct form of FSR is given by Eq. (\ref{friedel}). Whenever the scattering matrix has a rank greater than 2, the connection between FSR in Eq. (\ref{friedel}) and Eq. (\ref{two}) is not straight forward. However, we can make this connection for each partial density of states (PDOS) and is shown below. As an example, let us consider the Argand diagrams for the three prong potential shown in Figs. \ref{f7}(a) and \ref{f7}(c). None of the Argand diagrams (e.g. Figs. \ref{f7}(a), \ref{f7}(c)) are closed in the first Riemann surface. Let us, for example, consider the Argand diagram for $ t_{31} $ tracing a contour $ C' $ ($PDABQFARS$ shown in Fig. \ref{f7}(c)) as energy is varied from $ 0 $ to $ E_{1} $. We can show that (see Appendix A),
\begin{eqnarray}
\int_{C'}d\theta_{t_{31}}&\approx &2\pi\int_{0}^{E_1}\frac{\rho(3,1)(E)}{|t_{31}|^2}dE \label{thirteen}
\end{eqnarray}
Now we can again state that for any closed curve like $ ABQFA $ in Fig. \ref{f7}(c), Eq. (\ref{thirteen}) is exact. That is,
\begin{eqnarray}
\oint_{ABQFA}d\theta_{t_{31}}&= &2\pi\oint_{ABQFA}\frac{\rho(3,1)(E)}{|t_{31}|^2}dE \label{t1}
\end{eqnarray}
This statement can also be alternately justified as follows. For a closed contour, L.H.S of Eq. (\ref{t1}) is zero. Also for a closed contour R.H.S of Eq. (\ref{t1}) will be zero as shown in Appendix B for any $ \rho(\alpha,\beta) $ inside a closed contour integral. Note that in Appendix B the expression that we have used for $ \rho(\alpha,\beta) $ is given by R.H.S of Eq. (\ref{buttiker2a}). Hence it follows that R.H.S of Eq. (\ref{t1}) is also zero, where $ \rho(3,1) $ or any $ \rho(\alpha,\beta) $ is given by the R.H.S of Eq. (\ref{buttiker2a}). The arguments below although stated for $ \rho(3,1) $ is therefore true for any $ \rho(\alpha,\beta)(E) $ inside a closed contour integral. Now,
\begin{eqnarray}
\oint_{ABQFA}d\theta_{t_{31}}=\oint_{ABQFA}\frac{d\theta_{t_{31}}}{dE} dE=0 \label{t2}
\end{eqnarray}
\begin{figure}
\centering
{
{\includegraphics[width=10cm,keepaspectratio]{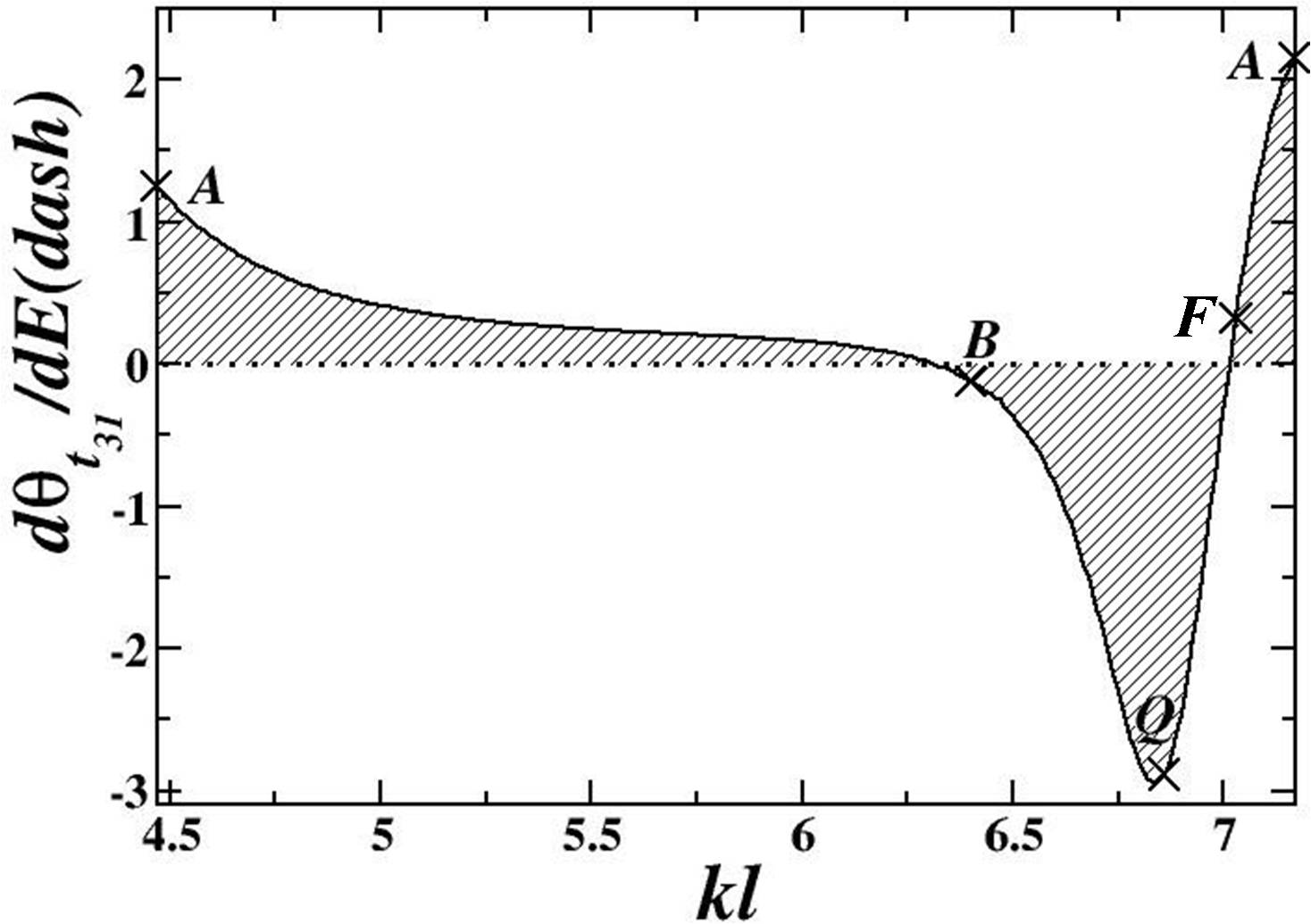}}
\caption{\label{fignew} $\frac{d \theta_{t_{31}}}{dE} $ as a function of $ kl $ for the three prong potential. Here $ l=1$, $ l_{2}=5 l$, $ e=1 $ and $ eVl= -1000 $.}
}
\end{figure}In Fig. \ref{fignew}, $ \frac{d\theta_{t_{31}}}{dE} $ is shown in the energy range covering the sub-loop $ ABQFA $ of Fig. \ref{f7}(c). As is implied by Eq. (\ref{t2}), $\frac{d\theta_{t_{31}}}{dE}$ is somewhere positive and somewhere negative to ensure the area under the curve (shaded region in Fig. \ref{fignew}) is zero. Similarly the R.H.S of Eq. (\ref{t1}) is zero implies that $ \frac{\rho(3,1)(E)}{|t_{31}|^2} $, will also be positive as well as negative in certain energy values (or $ kl $ values). Thus $ \rho(3,1) $ as given by the R.H.S of Eq. (\ref{buttiker2a}) is conclusively negative in some energy values. Although we have considered the there prong potential as an example the proof is valid for any potential whose Argand diagram shows a continuous phase drop due to a sub-loop as it is due to the fact that the sub-loop traces a closed contour. 

\begin{figure}
\centering
{
{\includegraphics[width=10cm,keepaspectratio]{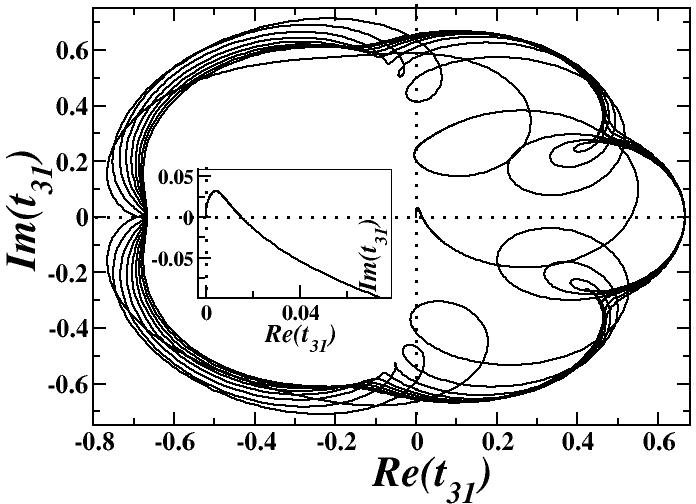}}
\caption{\label{fig8}Argand diagram for transmission amplitude $ t_{31} $, varying $ kl $ from $ 0 $ to $ 50 $, for the three prong potential. Here $ l=1$, $ l_{2}=5 l$, $ e=1 $ and $ eVl= -1000 $.}
}
\end{figure}

We will now show how negative slopes in scattering phase shift of mesoscopic systems are fundamentally different from that studied earlier \cite{landauer1} in 1D. Note the negative slope at point $ P $ in Fig. \ref{f7}(d). This kind of negative slope at very low energies can arise for scattering in 1D and one can easily check this for a square well potential. In terms of our analysis we understand the negative slope at point $ P $ in Fig. \ref{f7}(d) due to a convex trajectory at $ P $ in Fig. \ref{f7}(c) which is originating due to the fact that the Argand diagram starts from the origin and behaves anomalously as the trajectory starting from the origin is neither clockwise nor anti-clockwise with respect to the singular point (i.e. origin). See the expanded Argand diagram trajectory shown in the inset of Fig. \ref{f7}(c). The trajectory moves up, turns around and moves down to become convex in a small energy window and then winds around the origin anti-clockwise. Although $ {d\theta_{t_{31}}\over {dE}} $ is negative at $ P $ in Fig. \ref{f7}(d), there is no conclusive evidence that PDOS $ \rho(3,1) $ is negative at energy corresponding to point $ P $. Such a negative slope is fundamentally different from the negative slope at $ Q $ in Fig. \ref{f7}(d) that originate from a closed sub-loop $ABQFA$ in Fig. \ref{f7}(c), that we encounter only in Q1D and mesoscopic scattering. We have shown that the sub-loop seen in Fig. \ref{f7}(c) implies the presence of this negative slope in scattering phase shift $ \theta_{t_{31}} $ and also implies PDOS $ \rho(3,1) $ is negative. We have shown that an Argand diagram for such a scattering matrix element curls around and forms a sub-loop without violating the topological constraints of Eq. (\ref{two}). The line integration along the sub-loop $ ABQFA $ in Fig. \ref{f7}(c) does not contribute to the line integration over the trajectory $PDABQFARS$ of Fig. \ref{f7}(c) or $ I $ in Eq. (\ref{two}) is unaffected by the presence or absence of a sub-loop in the closed contour $ C $. Negative slopes of this second type (i.e. observed at $ Q $ in Fig. \ref{f7}(d)) that we have discussed here in-fact can appear or disappear very easily and can be found at much higher energies. In Fig. \ref{fig8}, we have shown the Argand diagram for $ t_{31} $ upto very high value of energy (or, $ kl $). We can see many sub-loops which again implies the presence of negative slopes in scattering phase shift and negative PDOS. Sometimes there is a cusp in the Argand diagram and such a cusp means a sub-loop has disappeared \cite{mic2}. Thus the phase drop will also disappear as we go to such energies. Disappearance of a sub-loop can be demonstrated by varying any other parameter like $ V, l, l_2 $ of the three prong geometry. 

$ \rho(3,1) $ being negative is a counter intuitive feature of quantum mechanics and can have interesting physical significance. Obviously, ac-response of a mesoscopic system will change drastically if $ \rho(\alpha,\beta) $ in Eq. (\ref{buttiker11}) changes sign. Also, it means an electron that is incident along lead 1 and transmitted to lead 3, dwells in some negative number of states (PDOS is negative) inside the scatterer. Total charge being electronic charge times number of states will be positive for these negatively charged electrons. So other electrons that are incident along lead 1 and transmitted to lead 2 or reflected back to lead 1 will be attracted by this positively behaving charge of electrons going from 1 to 3. This could be the explanation for the electron-electron attraction observed in the numerical experiment of ref. \cite{pramana}, where no explanation could be given. In the next section we will argue that this could have been also observed in an experiment.

Now let us try to understand if there is a general connection between negative slope and injectance becoming exact as observed in some earlier works \cite{satpathi} for specific potentials. 
Eq. (\ref{t1}) holds for the integrals and does not imply equality of the integrands. However, using Eq. (\ref{buttiker3}) we can write
\begin{eqnarray}
\rho(3,1)&\approx & \frac{1}{2\pi}|t_{31}|^2\frac{d\theta_{t_{31}}}{dE}
\label{sixteen}
\end{eqnarray}
\begin{eqnarray}
\rho(2,1)&\approx & \frac{1}{2\pi}|t_{21}|^2\frac{d\theta_{t_{21}}}{dE} 
\label{seventeen}\\
\rho(1,1)&\approx & \frac{1}{2\pi}|r_{11}|^2\frac{d\theta_{r_{11}}}{dE}
\label{eighteen}
\end{eqnarray}
Finally using Eqs. (\ref{sixteen}), (\ref{seventeen}) and (\ref{eighteen}), we can write
\begin{eqnarray}
\rho(1,1)+\rho(2,1)+\rho(3,1)\approx \nonumber\\
\frac{1}{2\pi}|r_{11}|^2 \frac{d\theta_{r_{11}}}{dE}+\frac{1}{2\pi}|t_{21}|^2\frac{d\theta_{t_{21}}}{dE}+\frac{1}{2\pi}|t_{31}|^2\frac{d\theta_{t_{31}}}{dE} \label{FSR}
\end{eqnarray}
L.H.S of Eq. (\ref{FSR}), i.e., $ \sum_{\alpha}\rho(\alpha,\beta) $ is well known as injectance as defined by the R.H.S in Eq. (\ref{buttiker4a}). The R.H.S of Eq. (\ref{FSR}) is the semi-classical limit of injectance. The topological interpretation of Eqs. (\ref{sixteen}), (\ref{seventeen}), (\ref{eighteen}) in terms of Argand diagrams leading to Eq. (\ref{FSR}) is very useful. The correction term to Eq. (\ref{FSR}) is known \cite{satpathi,lev} i.e.,
\begin{eqnarray}
\rho(1,1)+\rho(2,1)+\rho(3,1)= \nonumber\\
\frac{1}{2\pi}\left[ |r_{11}|^2 \frac{d\theta_{r_{11}}}{dE}+|t_{21}|^2\frac{d\theta_{t_{21}}}{dE}+|t_{31}|^2\frac{d\theta_{t_{31}}}{dE}+\frac{m^*|r_{11}|}{\hbar k^2}sin(\theta_{r_{11}})\right] 
 \label{FSR1}
\end{eqnarray}
\begin{figure}[h]
\centering
{{\includegraphics[width=\textwidth ,keepaspectratio]{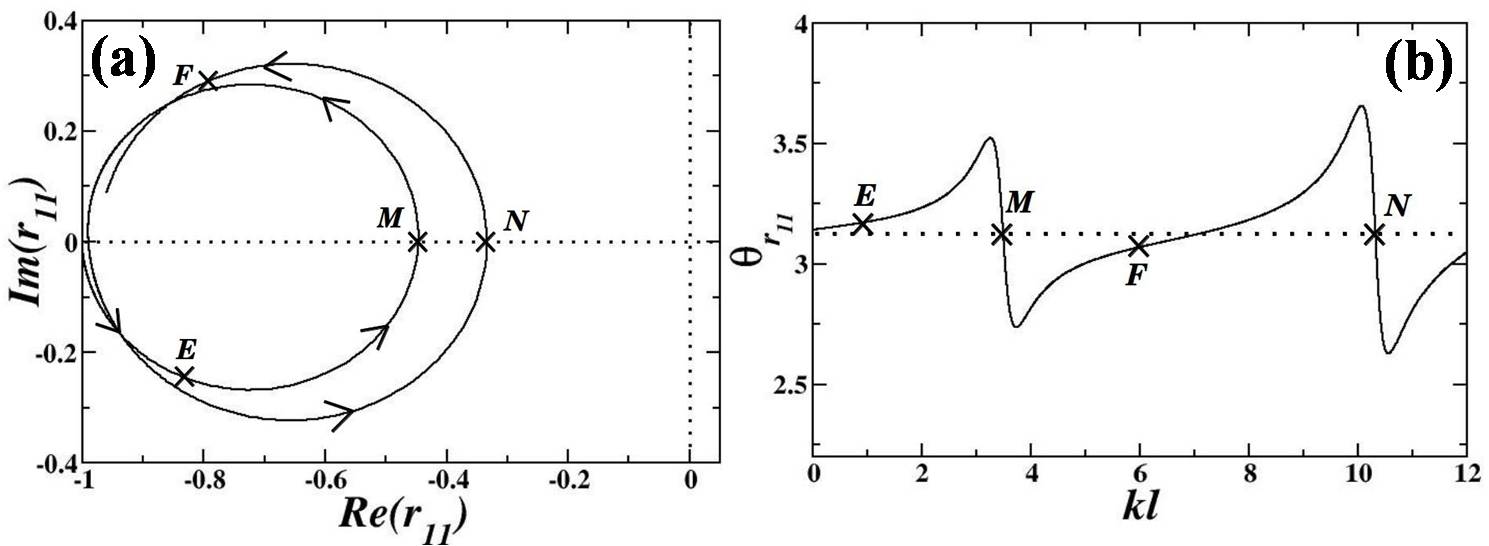}}}
\caption{\label{fig9new}(a) Argand diagram for reflection amplitude $ r_{11} $ and (b) plot of reflection  phase shift $ \theta_{r_{11}} $, versus $ kl $ for the three prong potential. Here $ l=1$, $ l_{2}=5 l$, $ e=1 $ and $ eVl= -10000 $.}
\end{figure}
\begin{figure}
\centering
{{\includegraphics[width=\textwidth ,keepaspectratio]{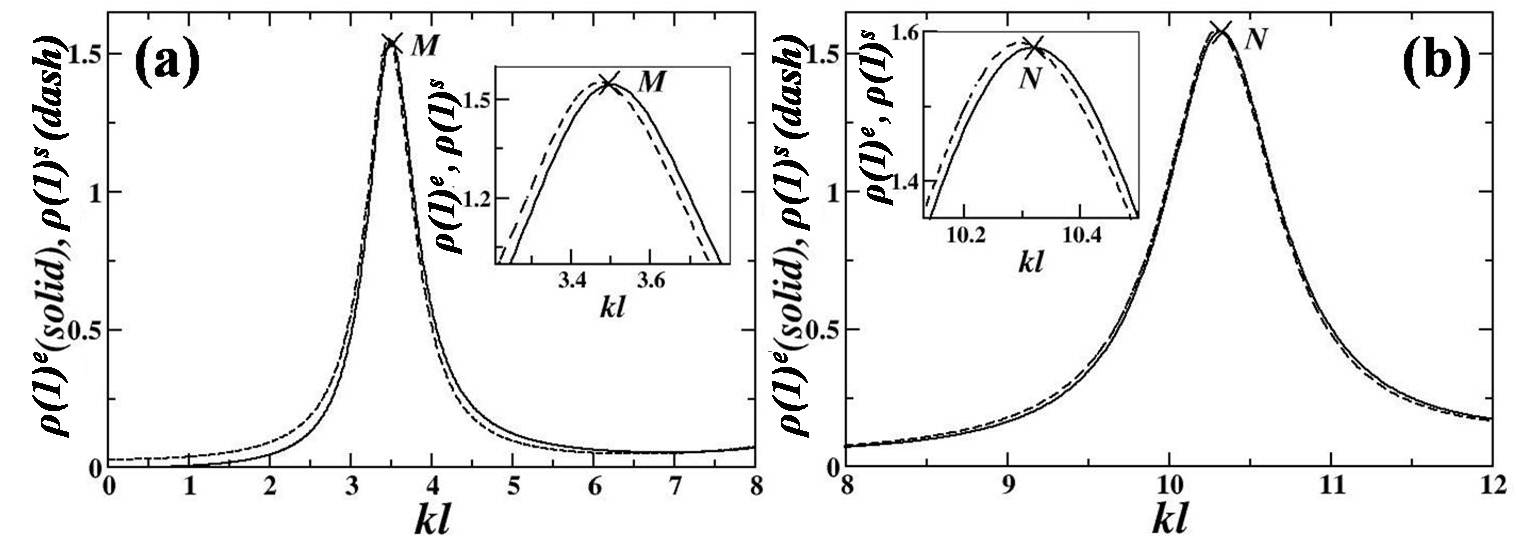}}}
\caption{\label{fig10new}Plot of exact injectance $ \rho(1)^{e} $ (solid line) and semi-classical injectance $ \rho(1)^{s} $ (dashed line) as a function of $ kl $ for the three prong potential. The peaks in the injectance are shown separately, (a) shows the first peak, (b) shows the second peak, for the same parameters as in Fig. \ref{fig9new}. The insets show the magnified curves at points $ M $ and $ N $.}
\end{figure} 
\begin{figure}[h]
\centering
{\includegraphics[width=.5\textwidth ,keepaspectratio]{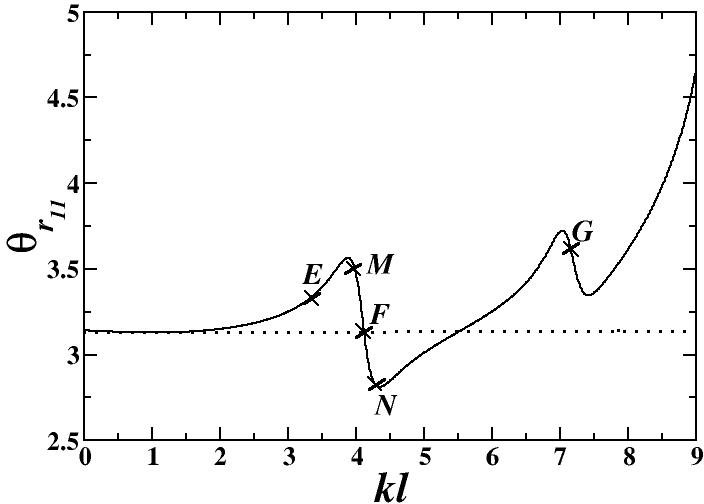}}
\caption{\label{fig9}Plot of reflection phase shift $ \theta_{r_{11}} $ versus $ kl $ for the three prong potential. Here $ l=1$, $ l_{2}=5 l$, $ e=1 $ and $ eVl= -1000 $.}
\end{figure}
\begin{figure}[h]
\centering
{{\includegraphics[width=\textwidth ,keepaspectratio]{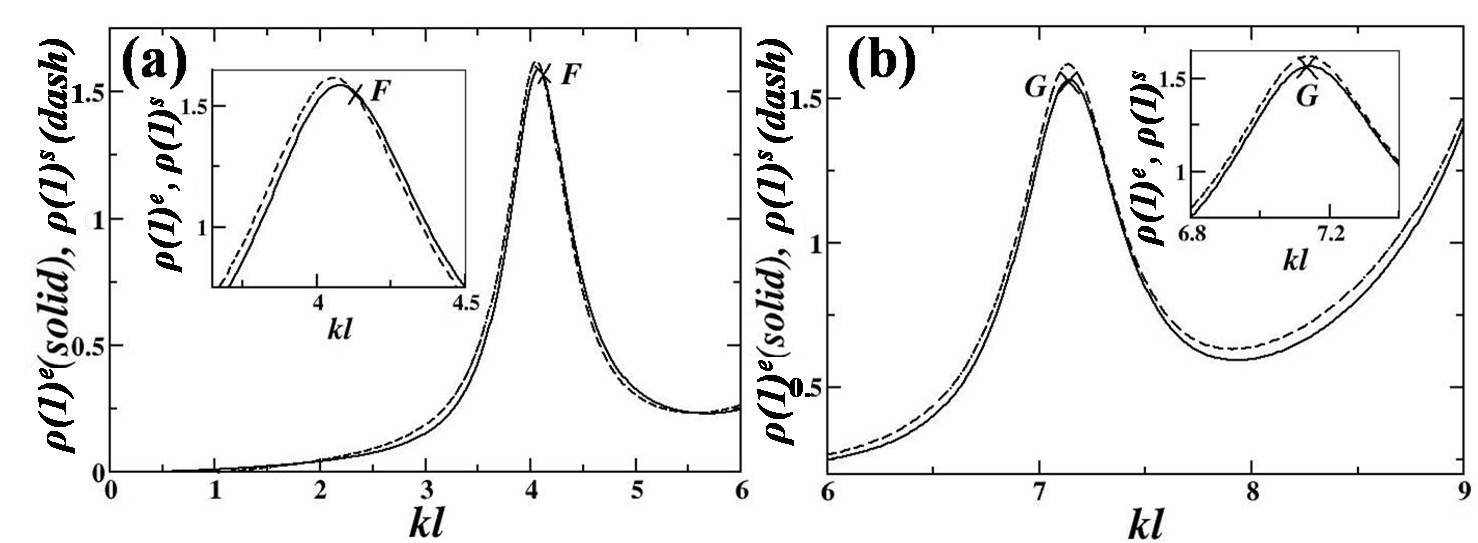}}}
\caption{\label{fig10}Plot of exact injectance $ \rho(1)^{e} $ (solid line) and semi-classical injectance $ \rho(1)^{s} $ (dashed line) as a function of $ kl $ for the three prong potential. The peaks in the injectance are shown separately, (a) shows the first peak, (b) shows the second peak, for the same parameters as in Fig. \ref{fig9}. The insets show the magnified curves at points $ F $ and $ G $.}
\end{figure}
Let us say that, 
\begin{eqnarray}
\rho(1)^{e} = \rho(1,1)+\rho(2,1)+\rho(3,1) 
\label{FSR2}
\end{eqnarray}
is exact injectance, defined by the R.H.S in Eq. (\ref{buttiker4a}). And,
\begin{eqnarray}
\rho(1)^{s} = \frac{1}{2\pi}\left[|r_{11}|^2 \frac{d\theta_{r_{11}}}{dE}+|t_{21}|^2\frac{d\theta_{t_{21}}}{dE}+|t_{31}|^2\frac{d\theta_{t_{31}}}{dE}\right]  \label{FSR3}
\end{eqnarray}
is generally referred to as semi-classical injectance. Eq. (\ref{FSR1}) implies that $ \rho(1)^{e} $ and $ \rho(1)^{s} $ will be equal at energies, where the correction term is zero or $  \vert r_{11} \vert sin(\theta_{r_{11}})=0 $. According to the arguments of Leavens and Aers \cite{lev}, in the semi-classical limit $ \vert r_{11} \vert\longrightarrow 0 $ and $ \rho(1)^{s}=\rho(1)^{e} $. But in the cases of studies on injectance or FSR in mesoscopic systems \cite{singlechannel,satpathi}, $ \vert r_{11} \vert\neq 0 $ at the energy where $ \rho(1)^{s}=\rho(1)^{e} $, but not very consistently as already discussed in the paragraph after Eq. (\ref{friedel}). We will show how the Argand diagram topology is responsible for this behaviour and therefore provides a general understanding. In Figs. \ref{fig9new}(a) and \ref{fig9new}(b), we have shown the Argand diagram and phase shift for scattering matrix element $ r_{11} $, respectively. $ kl $ is varied from 0 to 12 in both the plots. The Argand diagram in Fig. \ref{fig9new}(a) is restricted to one side of the phase singularity (i.e. origin) and in the first Riemann surface resulting in sub-loops. This will naturally mean that, the contour has both concave and convex parts in the trajectory. Scattering phase shift in Fig. \ref{fig9new}(b) increases with $ kl $, reaches a peak value and then drops to become $ \pi $ at $ M $. The pattern repeats as $ kl $ increases and the scattering phase shift becomes $ \pi $ again at $ N $. Therefore the correction term $   \vert r_{11} \vert sin(\theta_{r_{11}}) $ to semi-classical injectance (Eq. (\ref{FSR1})) is zero at $ M $ and $ N $. Thus at $ M $ and $ N $, the semi classical injectance (Eq. (\ref{FSR3})) is exact. The exactness of semi-classical injectance is shown in Fig. \ref{fig10new} (for clarity see the inset) at points $ M $ and $ N $ corresponding to same $ kl $ values as in Fig. \ref{fig9new}. For monotonously increasing phase, Argand diagram extends to higher Riemann surfaces and line integrals include the effect of other singularities in higher Riemann surface. Of course phase can be integral multiples of $ \pi $ (i.e., $ 2\pi, 3\pi, .. $), but for open Argand diagram trajectories as argued before, this is not a consistent behaviour, except in some simple one dimensional scattering problems. This inconsistent behaviour can be for example, checked for an one dimensional Aharanov-Bohm ring with different arm lengths. Drops in $ \theta_{r_{11}} $ resulting in $ \theta_{r_{11}} $ being $ \pi $ and $ sin(\theta_{r_{11}})=0 $, leading to semi classical injectance being exact can be understood from the Argand diagram in a single Riemann surface, involving a single phase singularity in the line integral and hence is not an accident. Drops in $ \theta_{r_{11}} $ is a pure quantum mechanical behaviour and hence exactness of semi classical injectance at the energies corresponding to the phase drops, is counter-intuitive. As argued before, these drops coming from sub-loops in Argand diagram are tunable and can be removed by varying some parameter. In Fig. \ref{fig9}, we plot the phase behaviour for the same system, after decreasing the potential $ V $ of the same system as in Figs. \ref{fig9new} and \ref{fig10new}. The phase drop at point $ F $ as usual decreases to $ \pi $, and hence semi classical injectance is exact at this point which can be seen in Fig. \ref{fig10}(a). Point $ F $ is marked at the same value of $ kl $ as in Fig. \ref{fig9}. But, in Fig. \ref{fig9}, the phase drop at $ G $ is now not sharp enough to decrease to $ \pi $. It is due to tuning the potential, that the sub-loop has now reduced in area, and consequently the drop has also reduced and will eventually disappear as $ V $ is decreased further. At such point $ \rho(1)^{e} $ is not equal to $\rho(1)^{s} $, i.e. semi classical injectance is not exact in spite of a phase drop. This can be seen in Fig. \ref{fig10}(b)) where point $ G $ is marked at the same value of $ kl $, as in Fig. \ref{fig9}. One would have expected that when we decrease the potential and make it weaker, semi-classical behaviour will be favoured. But on the contrary, for stronger potential the semi-classical injectance is exact at point $ N $ in Fig. \ref{fig9new}(b), while for a weaker potential in the same system the semi-classical injectance is not exact at point $ G $ in Fig. \ref{fig9}(b). Therefore, the drops in scattering phase shift of mesoscopic system, originating from sub-loops in Argand diagram involves completely new physics. One has to discard the usual concept of semi-classical regimes wherein the de-Broglie wavelength of the electron is much smaller than the scale of the potential, mathematically expressed by Eq. (\ref{buttiker16}) \cite{buttiker2}. 

 
\section{Comment on Experimental Observations}\label{exp}
\begin{figure}
\centering
{\includegraphics[width=\textwidth ,keepaspectratio]{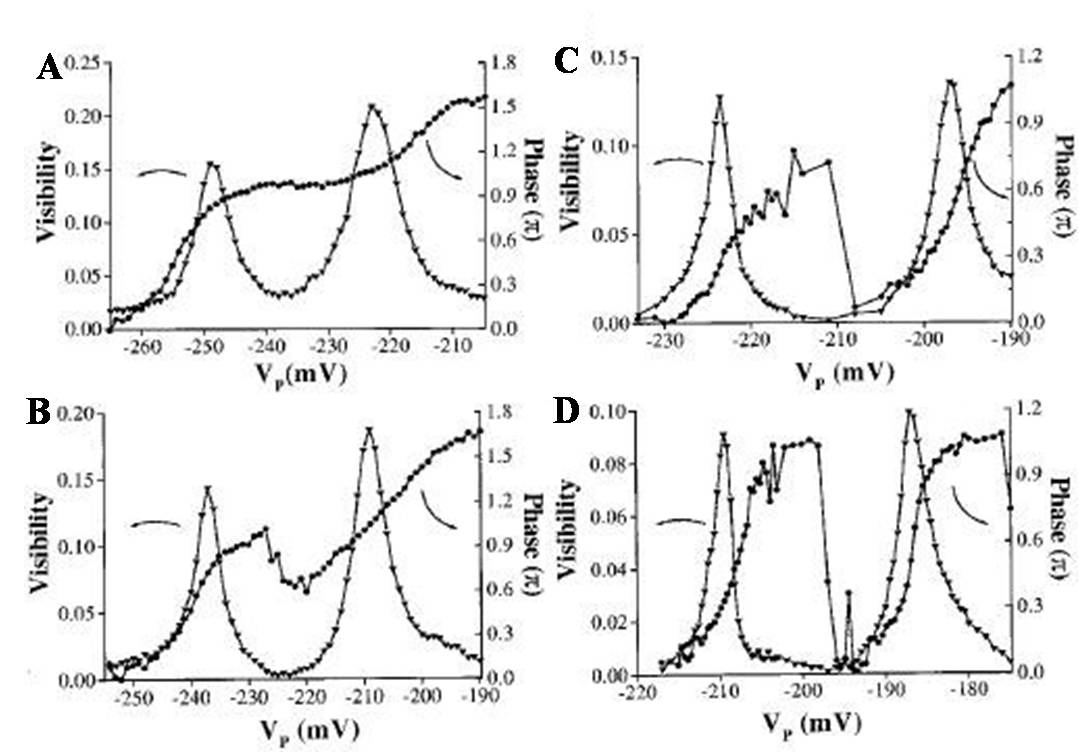}}
\caption[]{\label{fig 8}Transmission coefficient and scattering phase shift evolution in a QD as the coupling between the dot and leads is changed. This figure is taken with permission, from ref. \cite{yang}.}
\end{figure}
\begin{figure}
\centering
{\includegraphics[width=\textwidth ,keepaspectratio]{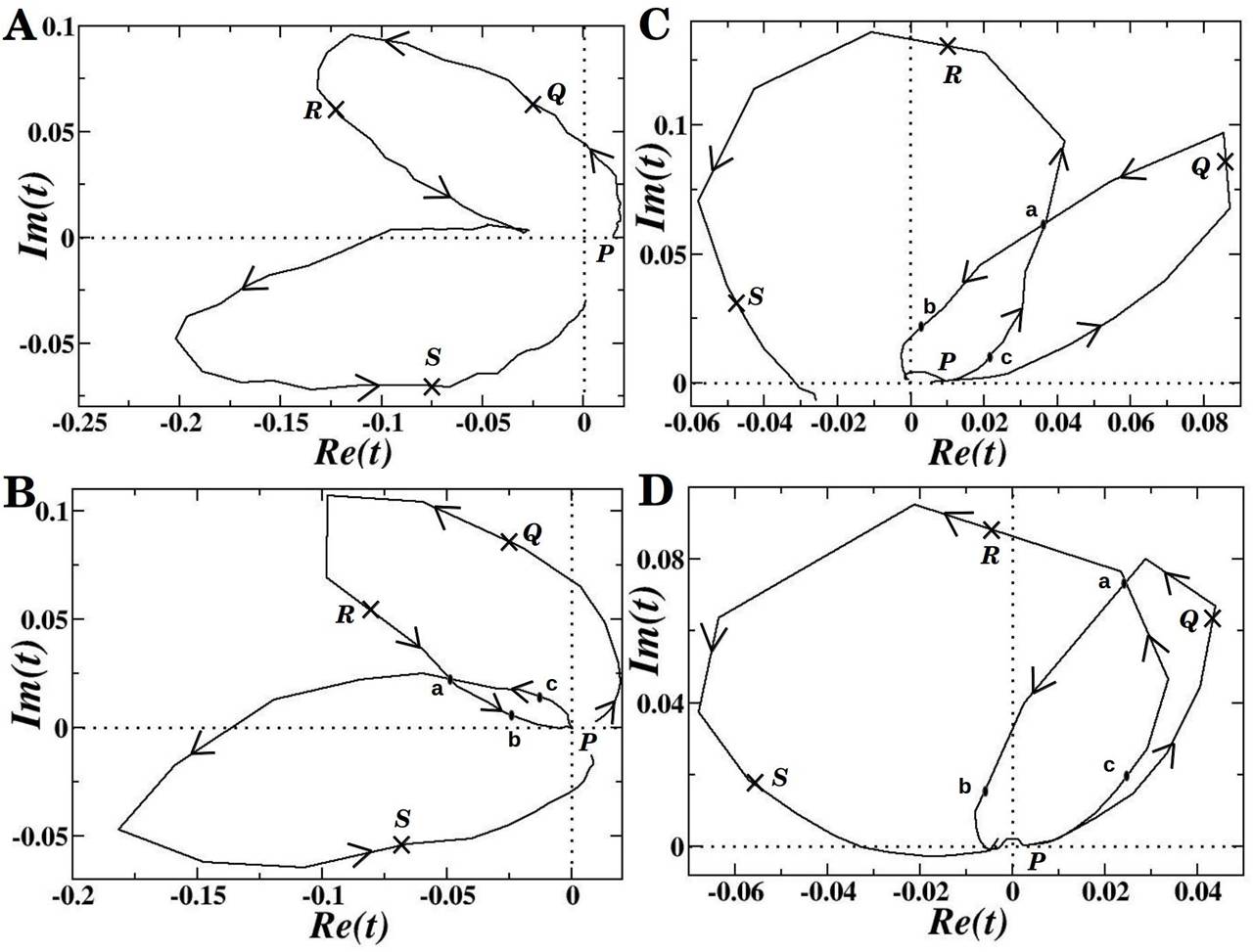}}
\caption[]{\label{fig 9}Argand diagram for the transmission amplitude, obtained from Fig. \ref{fig 8}. In this figure the labelling A,B,C,D follows that in Fig. \ref{fig 8}.}
\end{figure}
\begin{figure}  
\centering
{\includegraphics[width=\textwidth ,keepaspectratio]{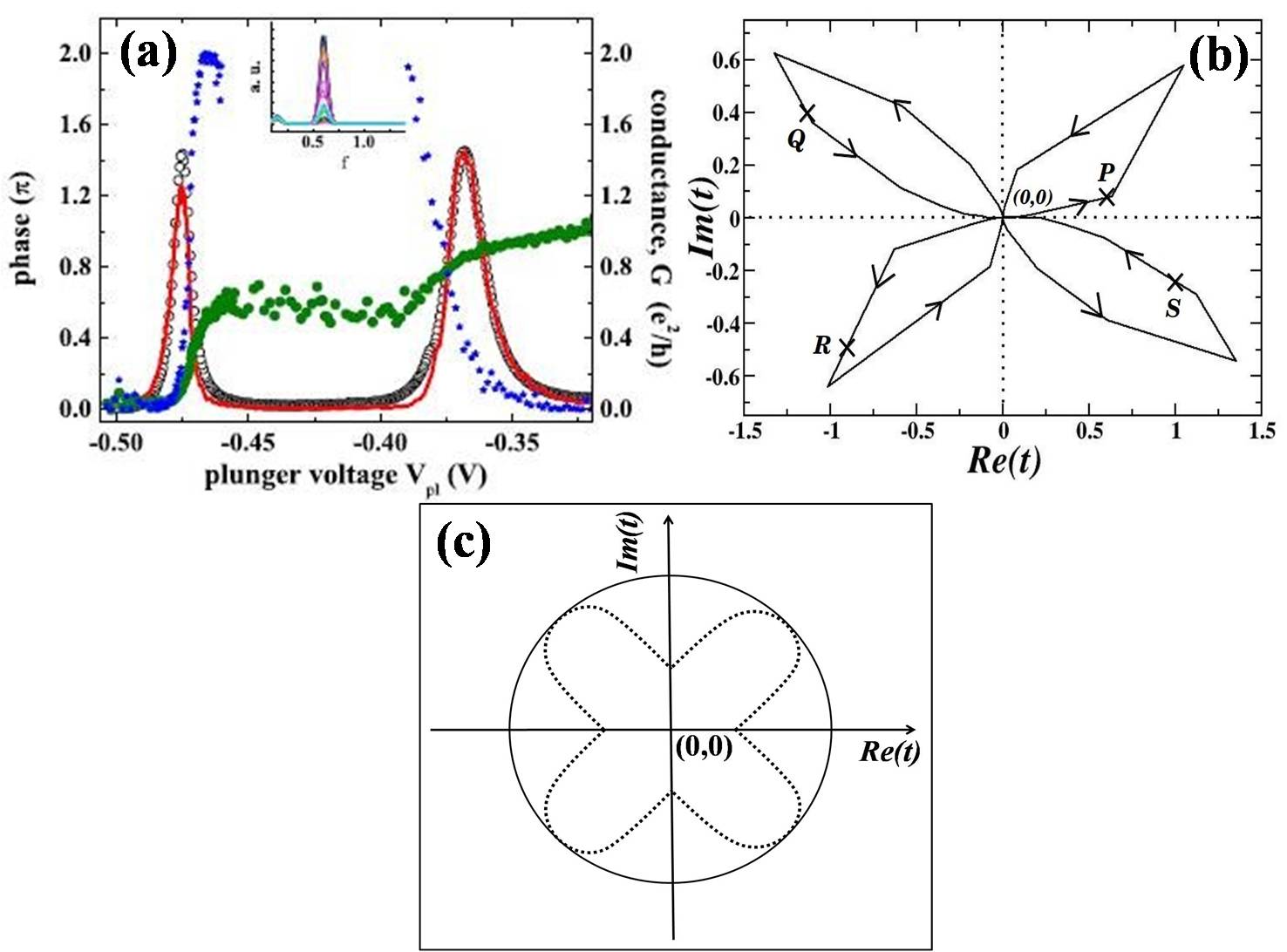}}
\caption[]{ \label{fig 10}(a) Typical transmission coefficient and scattering phase shift of a QD in the Kondo regime. This figure is taken with permission from ref. \cite{zaf}. (b) Argand diagram for transmission amplitude constructed from the experimental data. (c) We hope future experiments will give such Argand diagrams in greater resolution and how they evolve. For example an Argand diagram that encircles the phase singularity (solid line) can develop lobes (dotted line) and finally reduce to the Argand diagram in (b) as Kondo effect sets in.}
\end{figure}
We will try to construct the Argand diagram from the experimental data given in refs. \cite{yang,zaf}, to see if something can be understood.
Yang Ji et al. \cite{yang} measured the transmission coefficient and transmission phase shift of a quantum dot (QD). The experimentally observed transmission coefficient and transmission phase shift as a function of gate voltage, for different coupling strengths between QD and the leads, are shown in Fig. \ref{fig 8}. A phase drop is observed in Fig. \ref{fig 8}D when the coupling strength is small. As coupling gradually increases, the dot enters the strongly interacting regime and the phase drop decreases and finally vanishes. This gradual phase disappearance follows the sequence D to A in Fig. \ref{fig 8}. We have seen in our theoretical calculations that such phase drops can disappear and there are two possible reasons. One possibility is that when the coupling to leads is changed then it introduces some de-coherence for which the Argand diagram trajectory crosses the singularity and $ I $ changes from 0 to 1. The second possibility is that $ I $ remains conserved and the Argand diagram trajectory develops a sub-loop. Argand diagrams can be constructed from the experimental data given in Fig. \ref{fig 8} and is shown in Fig. \ref{fig 9}. There is a lot of fluctuations in the data probably due to experimental error. However, one can roughly see that there is a sub-loop $ abca $ in Fig. \ref{fig 9}, that gradually decreases in area as we go from D to A and clearly disappears in A with the formation of a cusp. The experimentalists had thought that as coupling increases, Kondo effect sets in and the disappearance of phase drop could be due to that. However, the phase behaviour seen in Fig. \ref{fig 8}A is not typical of Kondo resonance. Typical phase behaviour of Kondo resonance is that, it increases by $ \frac{\pi}{2} $ and forms a plateau. Such a behaviour was observed by the same experimental group \cite{zaf} subsequently, in which they restricted the dot occupancy to one electron, and is shown in Fig. \ref{fig 10}(a). One may conclude that the disappearing phase behaviour in Fig. \ref{fig 8}A has nothing to do with Kondo effect. We have already discussed in section~\ref{inj}, that when Argand diagram of transmission amplitude shows a sub-loop, PDOS associated with the transmitting electrons is negative. The transmitted particles and reflected particles can attract each other and that can result in a bound state. This bound state is expected to be effective only in the strongly interacting regime, i.e. for the case of Fig. \ref{fig 8}A. In the weakly interacting regime, we may not see any effect of this interaction induced bound state and the phase drops occur very generally in scattering by a mesoscopic system. We hope future experiments will focus on how Argand diagram evolves as one observes changes in scattering phase shift.
\section{Conclusions}
Argand diagram of scattering matrix elements are drawn for different model potentials and for a few experimental data. Several conclusions can be drawn from the topology of the Argand diagram without referring to the Hamiltonian or to the scattering potential. In 1D, 2D and 3D, the Argand diagram trajectory encircles the phase singularity. But in mesoscopic systems we find that the Argand diagram develops sub-loops. The sub-loop does not enclose the phase singularity at the origin and hence topologically allowed. The sub-loop can therefore appear or disappear on varying some parameter. It does not matter what parameter (say $ E $) is varied to obtain the Argand diagram or what parameter (say $ V $) is varied to make the sub-loop disappear. Many unexplained features so far can be explained by the appearance and disappearance of such a sub-loop. When the sub-loop appears there will be a gradual drop in the scattering phase shift and when the sub-loop disappear the drop will also disappear. Hence appearance and disappearance of phase drop is also very natural and poses no conceptual problem. Just as the sub-loop appears on varying some parameter, it can also grow in size as the parameter is varied. As the sub-loop becomes large and comes closer to the origin the phase drops also become large and sharp making the scattering phase shift decrease to $ \pi $ and then injectance or Friedel sum rule becomes exact. This is very counter intuitive as the strong phase drop signifies onset of pure quantum mechanical behaviour while Friedel sum rule is expected to become exact in semi-classical regimes. Also we prove that whenever there is a sub-loop (big or small) there will be negative partial density of states. For example if there is a sub-loop in the range $ \Delta E $ then there is also negative partial density of states in the range $ \Delta E $. Conclusive evidence of negative partial density of sates in a real system has never been reported before. Since all these results are drawn from the properties of the Argand diagram, the results are general and independent of the Hamiltonian or the scattering potential. The physics originating from sub-loops is completely new and upsets our way of understanding semi-classical behaviour.
\section{Acknowledgement}
One of us (PSD) would like to thank Prof. Sir M. V. Berry for useful discussions.
\renewcommand{\theequation}{A.\arabic{equation}}    
\setcounter{equation}{0}  
\section*{Appendix A}  \label{app1}
From Eq. (\ref{buttiker2a}), we can write
\begin{eqnarray}
\rho(3,1)&=& -\frac{1}{2\pi }\int_{sample} d\textbf{r}^{3} |t_{31}|^2\frac{\delta \theta_{t_{31}}}{\delta V(\textbf{r})} 
\label{fifteen}
\end{eqnarray}
In Eq. (\ref{fifteen}) using the well known semi classical approximation  $ \int_{sample} d\textbf{r}^{3}\frac{\delta t_{31}}{\delta V(\textbf{r})}\cong -\frac{dt_{31}}{dE} $ \cite{buttiker2}, we get, (see Eq. (\ref{buttiker3}))
\begin{eqnarray}
\rho(3,1)&\approx & \frac{1}{2\pi}|t_{31}|^2\frac{d\theta_{t_{31}}}{dE} \label{sixteenap}
\end{eqnarray}
Now, for a contour $ C' $ traced when energy is varied from $ 0 $ to $ E_{1} $,
\begin{eqnarray}
\int_{C'}d\theta_{t_{31}}&=& \int_{0}^{E_1}\frac{d\theta_{t_{31}}}{dE}dE \nonumber\\
&=& \int_{0}^{E_1} \frac{\frac{1}{2\pi}|t_{31}|^2\frac{d\theta_{t_{31}}}{dE}}{\frac{1}{2\pi}|t_{31}|^2}dE \nonumber \\
\text{Using Eq. (\ref{sixteenap})}\hspace{1cm}\int_{C'}d\theta_{t_{31}}&\approx& 2\pi\int_{0}^{E_1}\frac{\rho(3,1)}{|t_{31}|^2}dE
\label{thirteenap1}
\end{eqnarray} 
\renewcommand{\theequation}{B.\arabic{equation}}    
\setcounter{equation}{0}  
\section*{Appendix B}  \label{app2}
We know that in a scattering problem increasing incident energy by $ dE $ is equivalent to decreasing the potential globally by a constant amount $ \Delta\varepsilon $, such that $ dE=-e\Delta\varepsilon $, where $ e $ is particle charge that we will set to $ 1 $ to simplify our arguments. That is, the new potential is $ V'(\textbf{r})=V(\textbf{r})-\Delta\varepsilon $. Hence if we can generate a closed sub-loop in the Argand diagram by varying $ E $, then we can also do so by globally changing the potential and for such a closed contour like $ ABQFA $ in Fig. \ref{f7}(c), 
\begin{eqnarray}
\oint_{ABQFA} \delta\theta_{s_{\alpha\beta}}=0 \nonumber\\
\text{i.e.}\hspace{1cm}-\oint_{ABQFA}\int_{global}\frac{\delta\theta_{s_{\alpha\beta}}}{\delta V(\textbf{r})}\Delta\varepsilon d\textbf{r}^3=0 \label{B1}
\end{eqnarray} 
Now we replace the global integration over $ \textbf{r} $ by an integration over the sample or the scattering region only, since we have seen that it can be done in case of closed contours or inside an integration of the type $\oint_{C}$ in Eq. (\ref{B1}). This has already been discussed in Eq. (\ref{buttiker18}).
\begin{eqnarray}
\therefore\hspace{1cm}-\oint_{C}\int_{sample}\frac{\vert s_{\alpha\beta}\vert^{2}}{\vert s_{\alpha\beta}\vert^{2}}\frac{\delta\theta_{s_{\alpha\beta}}}{\delta V(\textbf{r})}\Delta\varepsilon d\textbf{r}^3&=&0\nonumber\\
\text{or,}\hspace{1cm}-\oint_{C}\frac{1}{\vert s_{\alpha\beta}\vert^{2}}\Delta\varepsilon\int_{sample}\vert s_{\alpha\beta}\vert^{2}\frac{\delta\theta_{s_{\alpha\beta}}}{\delta V(\textbf{r})}d\textbf{r}^3&=&0\nonumber\\
\text{or,}\hspace{1cm}2\pi\oint_{C}\frac{\rho(\alpha,\beta)}{\vert s_{\alpha\beta}\vert^{2}}\Delta\varepsilon &=&0\nonumber\hspace{1cm}\text{Using Eq. (\ref{buttiker2a})}
\end{eqnarray}
Therefore, the R.H.S of Eq. (\ref{t1}) is justified if an electronic charge is multiplied to the numerator.

\end{document}